 \documentclass[preprint]{aastex63}

\usepackage[flushleft]{threeparttable}

\received{June 5, 2020}
\revised{Oct. 10, 2020}
\accepted{Dec. 7, 2020}
\submitjournal{PSJ}

\shorttitle{Thermodynamically Governed Interior Models of Uranus and Neptune}
\shortauthors{Bailey and Stevenson}

\begin{document}

\title{Thermodynamically Governed Interior Models of Uranus and Neptune}

\author[0000-0002-4769-8253]{Elizabeth Bailey}
\affiliation{California Institute of Technology \\
1200 E. California Blvd\\
Pasadena, CA 91125}
\affiliation{University of California, Santa Cruz \\
1156 High Street\\
Santa Cruz, CA 95064}
\author{David J. Stevenson}
\affiliation{California Institute of Technology \\
1200 E. California Blvd\\
Pasadena, CA 91125}




\begin{abstract}
Interior models of Uranus and Neptune often assume discrete layers, but sharp interfaces are expected only if major constituents are immiscible. Diffuse interfaces could arise if accretion favored a central concentration of the least volatile constituents (also incidentally the most dense); compositional gradients arising in such a structure would likely inhibit convection. Currently, two lines of evidence suggest possible hydrogen-water immiscibility in ice giant interiors. The first arises from crude extrapolation of the experimental H$_2$-H$_2$O critical curve to $\sim 3$ GPa \citep{Bali13}. The data are obtained for an impure system containing silicates, though Uranus and Neptune could also be ``dirty.'' Current ab initio models disagree \citep{SoubMil}, though hydrogen and water are difficult to model from first-principles quantum mechanics with the necessary precision. The second argument for H$_2$-H$_2$O immiscibility in ice giants, outlined herein, invokes reasoning about the gravitational and magnetic fields. While consensus remains lacking, here we examine the immiscible case. Applying the resulting thermodynamic constraints, we find that Neptune models with envelopes containing a substantial water mole fraction, as much as $\chi_{\text{env}}' \gtrsim 0.1$ relative to hydrogen, can satisfy observations. In contrast, Uranus models appear to require $\chi_{\text{env}}' \lesssim 0.01$, potentially suggestive of fully demixed hydrogen and water. Enough gravitational potential energy would be available from gradual hydrogen-water demixing, to supply Neptune's present-day heatflow for roughly ten solar system lifetimes. Hydrogen-water demixing could slow Neptune's cooling rate by an order of magnitude; different hydrogen-water demixing states could account for the different heatflows of Uranus and Neptune.

\end{abstract}

\keywords{planets and satellites: interiors, planets and satellites: physical evolution}

\section{Introduction}\label{sec:intro}
Currently, Uranus and Neptune are the only planets in the solar system that still await visitation by an orbiter mission. Due to this relative lack of spacecraft coverage, as well as challenges to ground-based work resulting from their greater distance, knowledge about the so-called ice giants\footnote{This monicker assumes the intermediate density of these planets is due to a significant proportion of volatile species (i.e. ``ices'') in their interiors. However, as discussed in this work, there is actually no direct evidence that ices comprise a major proportion of these planets' mass. Observations of ices in the atmospheres do not necessarily inform the composition of the deep interiors, and the intermediate densities required to produce the mean densities and measured gravitational fields of these planets could, in principle, be produced in a scenario of mixed rock and hydrogen and no more methane than what is needed to explain the atmospheres.} is limited compared to the other solar system planets. But despite the general dearth of detailed information for Uranus and Neptune, the \textit{Voyager 2} flyby, as well as ongoing Earth-based observations, have revealed a clear paradox for these two planets, to be addressed in this work. Specifically, while Uranus and Neptune possess qualitatively similar magnetic fields--suggesting similar interior convective geometries distinct from all other dynamo-generating solar system bodies--these two planets simultaneously exhibit distinctly different intrinsic heat fluxes. A cohesive narrative has not yet been agreed upon to explain these similarities and differences between Uranus and Neptune.

The intrinsic heat fluxes of Uranus and Neptune have been determined by ground-based observations  (e.g. \cite{Fazioetal1976, Loewensteinetal1977a, Loewensteinetal1977b}), in conjunction with measurements from the infrared interferometer spectrometer (IRIS) on Voyager 2 \citep{Haneletal1986, conrath1989infrared,PearlCon1991}. These works have provided intrinsic heat flux measurements of Neptune that might be consistent with standard adiabatic cooling models \citep{NepBook} or suggestive of a source of excess luminosity \citep{Scheibeetal19}. In contrast, given current bond albedo estimates, the same cannot be said for measurements of Uranus's heat flux, which is an order of magnitude lower than Neptune's (and consistent with the heat flux actually being zero). Proposed reasons for Uranus' low heat flux have invoked either a low initial formation temperature, or some mechanism that inhibits convection in the interior and prevents heat from escaping efficiently (e.g. \citealt{podolak1991models, Net16, LeconteChab12, Podolak19}). Overall, a commonly considered possibility has been that Uranus' interior experiences a process that blocks heat from leaving, and that Neptune is not subject to, or is less affected by, this mechanism. In light of our comments about uncertain composition, it is worth remembering that any statements about the heat content of these planets is necessarily uncertain because of the large differences in thermodynamic parameters for the constituents, especially the very high specific heat for hydrogen relative to ice or rock. Still, the simultaneous similarity of the observed magnetic fields of Uranus and Neptune appears to present a paradox to the assumption that Uranus experiences deep inhibited convection, but that Neptune's heatflow results from a fully adiabatic interior condition.

The similar magnetic fields of Uranus and Neptune are qualitatively different from all other dynamo-generating solar system bodies. In both planets, the magnetic dipolar component is offset from the planet center (0.3 $R_{\text{U}}$, 0.55 $R_{\text{N}}$), mathematically equivalent to the large quadrupolar moment of these bodies, and considerably inclined to the spin pole (Uranus: $60^\circ$, Neptune: $47^\circ$). Although recent Juno observations have demonstrated asymmetry in the Jovian dipole \citep{moore2018complex}, the major quadrupole components of the magnetic fields of Uranus and Neptune reside in stark contrast to the generally dipole-dominated fields generated by all other solar system dynamos \citep{Nessetal1986,Nessetal1989,Connerneyetal1987,Connerneyetal1991}. The unusual field geometry of Uranus and Neptune has been reproduced with models in which the dynamo source region is a convecting thin shell surrounding a stably stratified fluid interior \citep{StanBlox04, StanBlox06}, as well as turbulent thick- and thin-shell models \citep{soderlund2013turbulent}. While the intriguing possibility for a thick-shell turbulent regime has not been ruled out in either planet, thin-shell dynamo models in particular do seem to agree with the potential explanation of Uranus' low heat flux as resulting from heat entrapment in the deep interior of Uranus, due to inhibited convection beneath the convecting shell. However, if Uranus' low heat flux is the result of deep inhibited convection, it is then necessary to explain why Neptune has a significantly greater intrinsic heat flux than Uranus, despite exhibiting a similar magnetic field. 

One proposed means to generate the key differences between Uranus and Neptune, has been giant impacts. The origins of the significant obliquities of these planets remains an open question--especially the $98^\circ$ spin axis tilt of Uranus, although Neptune's $30^\circ$ misalignment is also non-negligible--and a collisional origin of tilting has long been proposed \citep{Safronov1966,slattery1992giant, kegerreis2018consequences}. A major problem with a collisional origin of the tilt has been the need to explain the equatorial orientation of the orbits of the Uranian moons and rings; however, \cite{Morbyetal12} have found that a multiple-collision scenario allows for sufficiently gradual tilting that the proto-satellite disk can re-align with the planet. As an alternative explanation, Uranus gradually tilted as the result of a resonance between its orbit and precession of its spin axis \citep{BoueLaskar10}. Recently, \cite{reinhardt2020bifurcation} have suggested that an oblique giant impact to Uranus and a head-on collision to Neptune could account for the planets' obliquities and the differences between their satellite systems. Furthermore, they suggest that a head-on impact to Neptune could account for Neptune's less centrally condensed state relative to Uranus (inferred from rotation and gravity data), as well as the differences in heat flow between the planets. While the giant impact hypothesis represents an intriguing possible explanation for the differences in heat flow between the two planets, it is also worth considering other possible reasons for the origin of the disparity in heat flow between Uranus and Neptune.

Highlighting the importance of understanding the solar system's ice giants, it has often been suggested that Uranus and Neptune are possibly our best local analogues to the numerous observed exoplanets having masses and radii intermediate between those of terrestrial planets and gas giants. In fact, planets in an intermediate mass and radius range between gas giant planets and terrestrial planets are now understood to be an extremely common product of planet formation, at least at closer stello-centric distances \citep{Batalha13}. Although low detection sensitivity at host star separations beyond $\sim 10 \text{ au}$ has ensured that no perfect exoplanetary analogues to the solar system's ice giants have yet been found, the presence of a significant proportion of both light and heavy constituents in Uranus and Neptune makes them our most readily accessible laboratories for investigating the interactions of planetary constituents within intermediate-mass planets. 

Moreover, it is often suggested that the intermediate sizes of Uranus and Neptune are due to their status as ``cores" that failed to attain runaway accretion before the solar nebula dissipated, in the core accretion model for giant planet formation. However, despite the central role of Uranus and Neptune in understanding rates of planet formation in our own solar system, uncertainty about the composition and structure of their interiors remains a major obstacle to understanding the provenance and formation conditions of these planets-- and accordingly, their position within the greater narrative of planet formation in our solar system. Compared to gas giants and small bodies composed entirely of ice and rock, intermediate-density planets such as Uranus and Neptune suffer from a degeneracy in composition (e.g. \citealt{podolak1991models}). From observations of the gravity fields of Uranus and Neptune, it is established that the heavier elements must be concentrated toward the center, and surrounded by an envelope dominated by hydrogen and helium. However, in lieu of additional constraints, there is not a unique compositional profile which satisfies the measured properties of these planets.

\begin{figure*}
\centering
\includegraphics[width=\textwidth]{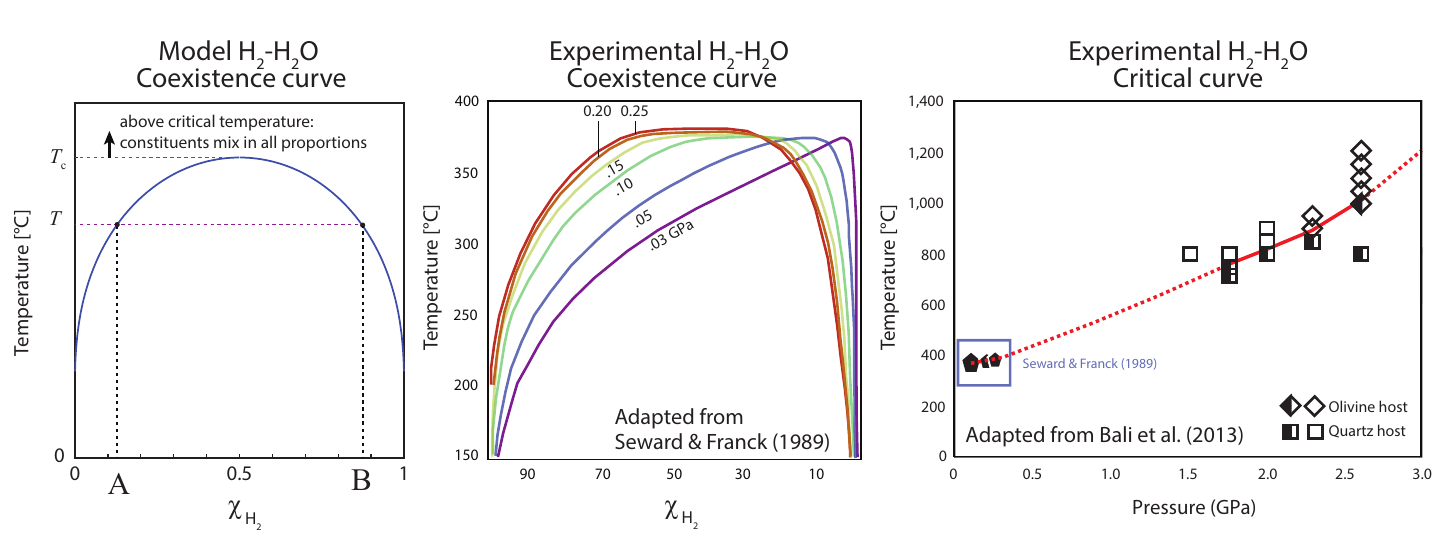}
\caption{Experimental data for the H$_2$-H$_2$O system, and a diagram showing a model coexistence curve and its relationship to the critical temperature $T_{c}$. For the purposes of this work, the ``critical curve'' refers to the critical temperature as a function of pressure. \textit{Left:} The peak of the coexistence curve occurs at the critical temperature $T_{c}$, above which the two species mix freely in any proportion. Below the critical temperature, the coexistence curve dictates the saturation compositions for coexisting phases. \textit{Center:} The coexistence curves and critical temperature have been determined up to $0.25$ GPa by \cite{SewFra81}, showing a trend toward increasing symmetry with pressure. \textit{Right:} The critical curve has been experimentally derived up to $\sim 3$ GPa by \cite{Bali13}, showing a roughly linear trend. The pentagonal markers show the critical temperature found by \cite{SewFra81}, while the square/diamond markers show the data found by \cite{Bali13}; black and white points indicate H$_2$-H$_2$O immiscibility. }
\label{fig:expmot}
\end{figure*}

\begin{figure}
\centering
\includegraphics[width=0.4\textwidth]{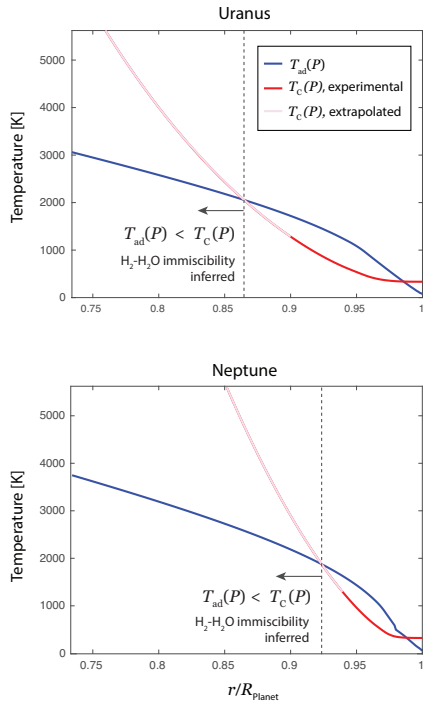}
\caption{Comparison of the experimental critical temperature (red) derived by \cite{Bali13} (Figure \ref{fig:expmot}), and its approximate extrapolation (pink) linear in pressure to beyond the $\sim 3$ GPa experimental limit, versus the adiabatic temperature profile (blue) in the outermost, hydrogen-dominant shell in models of Uranus and Neptune. A deep region of hydrogen-water immiscibility ($T_{\text{ad}}(P)<T_c(P)$) is predicted in the deeper regions of this layer, indicating a plausible phase transition in the interiors of these planets, although further laboratory data is warranted. While these temperature profiles refer to the best-fit Uranus and Neptune models discussed later in the text, from this rough extrapolation, a deep interior region of immiscibility is suggested for all compositions of the $H_{2}$-dominant shell considered in this work.}
\label{fig:RAT}
\end{figure}

Over the span of decades, numerous models have been constructed that satisfy the mass and observed gravity harmonics (up to $J_4$) of Uranus and Neptune. A traditional approach is to include several discrete, layers of uniform composition, with each layer typically composed primarily of ``gas," ``ices," and ``rock." These three terms refer to composition rather than the phase in which these materials occur: ``gas" refers to a solar-composition mixture, ``ice" refers to volatile hydrides such as H$_2$O, NH$_{3}$, and CH$_4$, while ``rock" generally refers to a combination of silicates and iron. In published works that invoke layers, their choice is not generally motivated by a specific physical rationale. Moreover, models which satisfy the inferred transition, from the hydrogen-dominated envelope to the denser mantle, using a substantial density gradient rather than discrete layers cannot satisfy the constraint that the dynamo magnetic fields require a well-mixed layer of sufficiently large radial extent. However, discrete interior layers (as opposed to a compositional gradient) are only expected to be stable when immiscible phases are present. In the case of terrestrial planets, formation of an iron core with a discrete core-mantle boundary occurs due to the immiscibility of iron in silicates at the relevant pressures and temperatures. In contrast, in the deep interiors of Jovian planets, it is expected \citep{WilsonMilitzer12a,WM12b} that dissolution of ice and rock in metallic hydrogen is thermodynamically favorable; therefore, if a core is present today, it is widely expected to be in the process of dissolving \citep{wahl2017comparing, debras2019new}. These are just a few examples of how mixing properties of major constituents govern the structure of planetary interiors. Another well-known case--helium immiscibility in giant planets, leading to helium rainout--will be discussed in a later section. This work addresses the implications of hydrogen-water immiscibility on the interiors of Uranus and Neptune.

The remainder of the paper is outlined as follows. In Section \ref{sec:method}, the method for constructing interior models is outlined. The results of this modeling effort are presented in Section \ref{sec:results}, while the implications are discussed in Section \ref{sec:disc}, followed by some concluding remarks in Section \ref{sec:conc}.

\section{Methods}\label{sec:method}

Multiple approaches have been used to produce static models of ice giant interiors. In the most typical overall approach to modeling these planets (e.g. \citealt{Pod76, HubMacfar80, Podolaketal1985, NepBook, podolak1991models, Podolaketal1995,Nettelmannetal13, Net16}), the details of the number and composition of interior layers are defined initially, and a density profile is then derived using equations of state of the chosen layer constituents. In an alternative set of approaches developed for these planets, density profiles are generated to satisfy the gravity harmonics without any a priori assumption of the composition or equations of state (e.g. \citealt{Marleyetal1995,Podolaketal00,Helledetal10}). 

Although the latter approach circumvents the need to adopt equations of state at thermodynamic conditions that are a major challenge to statically produce in the laboratory, the generated density profiles are not guaranteed to represent any physical mixture of plausible planetary constituents. Because in this work, we aim to constrain the space of possible layer compositions of Uranus and Neptune, we take the more traditional approach, by pre-defining the layers and their compositions.

In Uranus and Neptune, water is generally assumed to be the primary major constituent by mass. To satisfy the gravity harmonics, an underlying mantle with a density comparable to water, extending to $\sim 70$ percent of the total planet radius, must be overlain by a hydrogen-rich envelope. Therefore, in this work, hydrogen and water are explored as possible dominant constituents whose mixing properties might dictate the state of ice giant interiors. As discussed herein, constraints on H$_2$-H$_2$O miscibility remain to be fully characterized. However, some advances have been made in understanding the hydrogen-water system at conditions relevant to the interiors of ice giants, hence motivating this work. Figure \ref{fig:expmot} illustrates the existing experimental constraints for hydrogen-water mixing. In particular, two related thermodynamic curves that describe the mixing properties of hydrogen and water are the coexistence curve and critical curve. Figure \ref{fig:expmot} also shows a model coexistence curve for purposes of illustration.

Colloquially, it is often said that two immiscible species do not mix. However, it is more accurate to state that immiscible species do mix, but to a limited extent, in proportions specified by the coexistence curve for the binary system. This thermodynamic curve defines the compositions (typically in terms of mole fraction) at which minima of the Gibbs free energy of mixing occur, for a given pressure and temperature. Experimental and theoretical examples of the coexistence curve are shown in Figure 1. The pressure-dependent critical temperature $T_c$ is the maximum of the coexistence curve. The H$_2$-H$_2$O coexistence curve has been determined to $0.25$ GPa by \cite{SewFra81}. Because this experimental work does not attain conditions of the deep interiors of Uranus and Neptune, we employ a model coexistence curve, which will be discussed shortly in further detail, for the purposes of this work.

While the H$_2$-H$_2$O coexistence curve has long been known at pressures relevant to giant planet atmospheres, more recently, \cite{Bali13} have experimentally derived the critical temperature $T_c$ for the H$_2$-H$_2$O binary system up to $\sim 3$ GPa, a pressure range relevant to the deep interiors of Uranus and Neptune. Rough linear extrapolation of their result (Figure \ref{fig:RAT}) appears to suggest that the temperature deep in the ice giants may be below the H$_2$-H$_2$O critical temperature at those pressures, appearing to suggest possible immiscibility of hydrogen and water, and hence separated phases. However, a few major caveats arise when extrapolating the critical curve to higher pressures. First, while these experimental data suggest a roughly linear trend of the critical curve within the experimental range, there is no reason to expect the critical curve to continue linearly in pressure. In fact, to the contrary, as hydrogen approaches a more metallic state, the critical curve is expected to turn over (although the pressure at which this turnover begins, as well as its specific shape, are not known in detail). Indeed, \cite{wilson2011solubility} report solubility of water in hydrogen once 10-megabar pressures are reached. Moreover, \cite{SoubMil} reported results of ab-initio simulations which appear to possibly contradict H$_2$-H$_2$O immiscibility in the deep interiors of Uranus and Neptune. More specifically, they did not find evidence of concavity of the Gibbs free energy of mixing $\Delta G$ as a function of composition--in apparent contradiction to the experimental findings of \cite{Bali13}. In response to this discrepancy, \cite{SoubMil} argue that the experimental result may be due to contamination by the carrier silicates used in the experiment. On the other hand, an experiment contaminated by silicates may actually be more representative of the interiors of Uranus and Neptune than one that is not, as these planets likely include silicates as well. Clearly, more work is necessary to resolve the question of hydrogen-water miscibility in Uranus and Neptune. For the purposes of this work, we do not intend to make assertions about this question; however, we do explore the implications that hydrogen-water immiscibility would have for the interior states of these planets.

The discussion now turns from the critical curve to the specifics of the coexistence curve for hydrogen and water. While \cite{Bali13} have experimentally determined the critical temperature as a function of pressure to $3$ GPa, this finding does not inform the actual proportions at which hydrogen and water would be expected to mix at specific conditions. As mentioned previously, the specific compositions of coexisting (saturated) equilibrium phases are governed by the pressure-dependent coexistence curve, as was found experimentally by \cite{SewFra81} at pressures relevant to the atmospheres of these planets. As is evident in Figure \ref{fig:expmot}, at the lower pressures investigated in the experimental work, the coexisting phases are asymmetric, trending toward symmetric with increasing pressure. The low-pressure asymmetry can be attributed to the significant repulsion experienced by a (nonpolar) H$_2$ molecule when inserted into water, whereas the analogous effect in the H$_2$-dominant phase is lacking due to the greater (i.e. gas-like) intermolecular spacing of H$_2$ at low pressures. With increasing pressure, the hydrogen-rich phase becomes more closely packed, and the coexistence curve becomes increasingly symmetric, as shown by the experimental curves of \cite{SewFra81}. This symmetry behavior of binary phase diagrams at high pressure is common, and can be described in terms of the following simple model for a two-component regular solution:

\begin{equation}
\Delta G = \chi(1-\chi) \Delta E + kT(\chi \ln \chi + (1-\chi)\ln (1 - \chi))
\label{eq:chis}
\end{equation}

where $\Delta G$ is the Gibbs energy of mixing, $\Delta E$ is an interaction parameter, $\chi$ is the mole fraction of one of the components, $k$ is Boltzmann's constant, and $T$ is the temperature. As the system will move toward the state with the lowest available Gibbs energy, the equilibrium composition(s) is/are associated with minima in $\Delta G$. The critical temperature $T_c$, above which the two components mix in all proportions, is determined by the temperature above which $d^2\Delta G/dx^2$ is never negative. Below the critical temperature\footnote{Remember that $T_{c}$ depends on pressure.}, there are two solutions symmetric about $\chi = 1/2$. At and above the critical temperature, they collapse to one solution $\chi=1/2$, and phase separation does not occur for the case $T\geq T_{c}$. Taking this into account, for the binary system, it is possible to define the coexistence temperature $T_{\text{co}}(P)$ as the temperature for a given pressure at which a phase containing a water mole fraction $\chi<0.5$ of water coexists with a phase a water mole fraction $(1-\chi)$:

\begin{equation}
\frac{T_{\text{co}}(P)}{T_{c}(P)} = \frac{2(1-2\chi)}{\ln\big(\frac{1-\chi}{\chi} \big)}.
\end{equation}

However, because only extrapolatory inference of $T_{c}$ exists above $3$ GPa, the equilibrium compositions deep in these planets is not known, and numerous pairs of complementary (symmetric) compositions are therefore considered in this work. We cannot be sure that the phase diagram is symmetric at high pressure, since the two species are not similar in size or behavior, but a more nearly symmetric behavior is often observed in systems of two condensed (i.e., fluid density) phases exhibiting immiscibility \citep{Bernabeetal1987}. Asymmetry can be due either to difference in size between the two species or to different spacings, as in the case of a liquid and a vapor \citep{DamayLeclercq1991}. The experimental curves of \cite{SewFra81} show symmetry increasing with pressure, due to decreased difference in spacing between hydrogen versus water. However, the difference in molecular size between water and hydrogen suggests that a degree of asymmetry of the coexistence curve may be indicated at higher temperatures. We therefore allow coexisting equilibrium phase compositions to deviate from symmetric values by a factor of up to two.

\begin{figure}
\centering
\includegraphics[width=\textwidth]{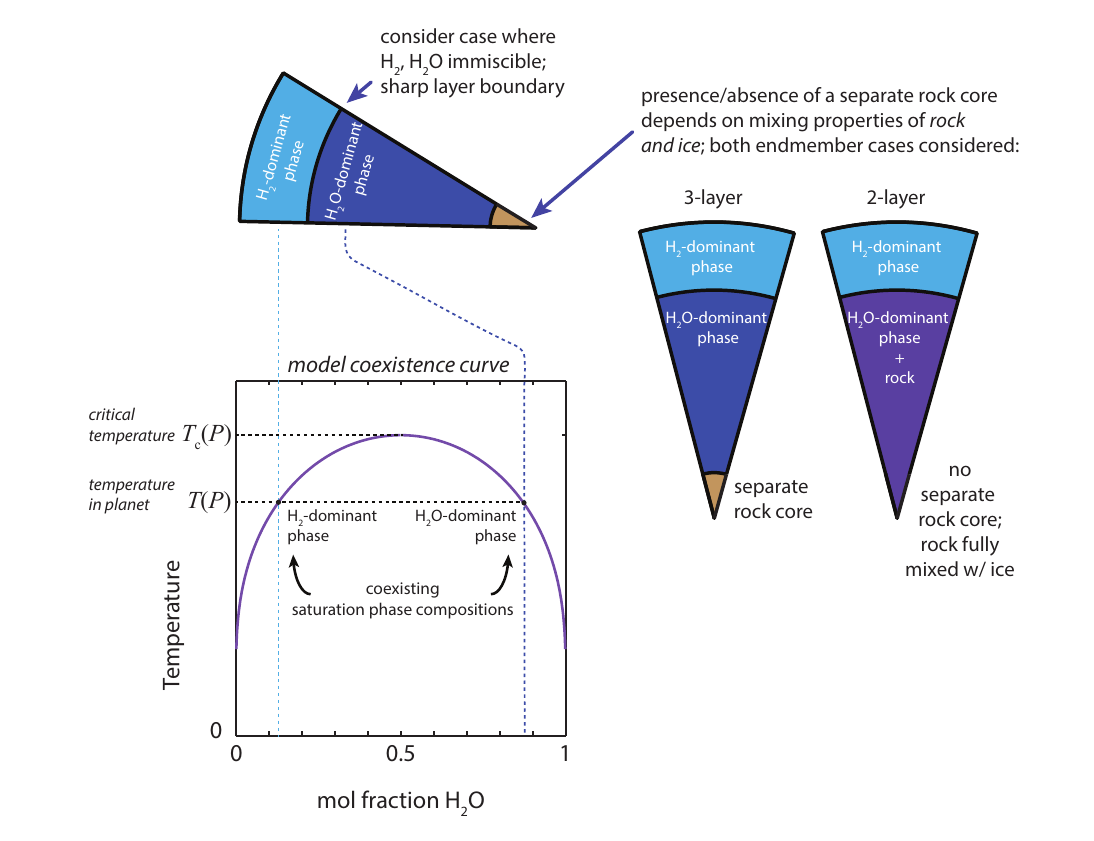}
\caption{Schematic diagram illustrating the models described in this work. We consider models that include a compositional discontinuity between the water-dominant and hydrogen-dominant layers, a structure expected to be stable only if hydrogen and water are immiscible. The plotted curve is the model coexistence curve (see also Figure \ref{fig:expmot}), with dotted lines tying the coexisting phase compositions for a given temperature and pressure to the hydrogen-dominant and water-dominant layers in the model. The inferred approximate symmetry of this hydrogen-water coexistence curve at interior pressures, in accordance with Equation \ref{eq:chis}, provides the novel thermodynamic constraint put forth in this work. This work adopts the assumption that water and other ices are enriched in the interior, and addresses the constraints on interior layer composition that would be obtainable if hydrogen and water are immiscible in the interiors. All models therefore contain separate layers of hydrogen-dominant and water-dominant composition. Models both with a separate rock core, and with homogeneous mixing of rock into the ice-dominant layer, are considered.}
\label{fig:scheme}
\end{figure}

Taking the symmetric rationale into account, we construct models of Uranus and Neptune, applying the constraint that models must be compatible with the coexisting equilibrium compositions implied by hydrogen-water coexistence diagram. In addition, the models must not be at odds with the observed magnetic fields of Uranus and Neptune--this rules out substantial compositional gradients in at least the outer $\sim 20\%$ of the planets, as such gradients would preclude the large-scale vertical motions necessary for a dynamo. Below the water cloud decks in Uranus and Neptune at tens or hundreds of bars, the thermodynamically permissible phases are thus either a water-rich ocean extending deep into the planet, or a hydrogen-dominant phase. Measurements of the gravitational moments, however, are at odds with the former, as they would imply too great a value of $J_{2}$. Therefore, we assume a hydrogen-dominant phase is present immediately below the water cloud decks of these planets. A schematic illustration of the thermodynamic constraint applied to this model, with the model critical curve described above, is shown in Figure \ref{fig:scheme}. Because they introduce unphysical layering, most published layered models either violate these requirements, or are in danger of doing so.

The models presented in this work are constructed and analyzed in three steps. First, layer compositions are chosen with $\chi$ and $(1-\chi)$ roughly symmetric about $0.5$, in agreement with the rationale discussed in Section \ref{sec:intro}. Next, taking into account these chosen layer compositions, density profiles are derived that satisfy each planet's radius and mean density. Finally, we apply a theory-of-figures approach (concentric Maclaurin spheroids; \citealt{Hubbard13}) to derive the gravity harmonics for each density profile, comparing them to observational constraints on the gravity field. The relevant observational constraints for Uranus and Neptune are given in Table \ref{tab:constr}, and are discussed in further detail in the following sections. 

\subsection{Choice and composition of model layers}

We now discuss the nature of the layers included in our static models. Models were considered which included both two and three layers (Figure \ref{fig:scheme}). In both cases, the outermost layer, the hydrogen-dominant ``envelope,'' consisted primarily of hydrogen and helium, with the proportion of water varied, and the proportion of atmospheric methane varied within observational bounds. The second layer, the ``mantle,'' consisted primarily of water, with ammonia and methane included in fixed amount relative to water, and the proportion of hydrogen included in this layer was varied. In the three-layer models, a separate core of silicates and iron was included. In a manner analogous to the case of hydrogen and water (the focus of this work), a sharp boundary between the core and mantle is expected only if rock and water are immiscible at the relevant conditions in these planets. However, the mixing properties of rock and ices at deep interior conditions are not known. Accordingly, two-layer models, lacking a separate rock core, were constructed based on the three-layer models, by taking the proportion of rock in the cores of the three-layer models, assuming the rock is mixed with the overlying ices and hydrogen.

A range of water-hydrogen ratios for the envelope, and associated hydrogen-water ratios for the mantle, are tested. Three-layer models are constructed with a range of mantle-envelope transition levels. Specifically, for each set of chosen layer compositions, the range of mantle-envelope transition depths is found for which it is feasible to construct a model satisfying the planet radius and mean density. Every feasible mantle-envelope transition depth has an associated rock core extent which permits the model to satisfy these basic constraints.

Having constructed this suite of three-layer models, associated two-layer models with mixed rock-ice mantles, and \textit{lacking separate rock cores}, are then constructed. The mantles in these models retain the same relative proportions of ices and hydrogen in the mantle, and the mantle is given an ice-rock ratio equivalent to the ratio collectively present in the mantle and core in the original three-layer model.

In this work, following the thermodynamic rationale discussed in Section \ref{sec:intro}, complementary molar ratios of H$_2$O to H$_2$ were considered in the gas-rich and ice-rich layers. That is, for each model, the molar ratio $\chi'_{\text{env}} \equiv \chi_{\text{H}_2\text{O}}/(\chi_{\text{H}_2\text{O}}+\chi_{\text{H}_2})$ of H$_2$O to H$_2$ in the hydrogen-rich envelope was assumed to be equivalent to the molar ratio $\chi'_{\text{man}} \equiv \chi_{\text{H}_2}/(\chi_{\text{H}_2\text{O}}+\chi_{\text{H}_2})$ of H$_2$ to H$_2$O in the underlying water-rich mantle. Moreover, as discussed in the previous section, to account for the possibility of an asymmetric binodal curve, models were also constructed in which these two ratios varied from one another by a factor of $2$. While the ratio of hydrogen and water in the adjacent layers is assumed to be thermodynamically governed, the fractions of other constituents--as well as small corrections to this assumption--are now discussed in detail. 

In the gas-rich layer, in addition to water and hydrogen, additional constituents are expected to be present, most notably Helium, CH$_4$, and NH$_3$. A solar proportion of He relative to diatomic hydrogen was assumed. Ammonia abundances in the envelope were chosen according to the atmospheric values given in \cite{LoddersFegley1994}--however, especially as NH$_3$ is expected to be depleted by interaction with H$_2$S \citep{dePateretal1991}, these values might not represent ammonia abundances further down in the envelope. As discussed later in this work, due to its polar nature, ammonia may plausibly mix preferentially with water. It should be noted that, for the purposes of understanding the density profiles and gravity harmonics of these planets, ammonia and water are essentially interchangeable. For the purposes of this paper, we are motivated by the laboratory result of \cite{Bali13} to focus on the possible effect of hydrogen-water demixing on ice giant structure and evolution, but the reader should keep in mind the uncertainty of the ammonia composition in the envelope and its potentially interchangeable role with water in the model framework put forth in this work. 

Moreover, the observed atmospheric methane abundances of Uranus and Neptune, relative to H$_2$ ($n/\text{H}_2$ = $0.023 \pm0.006$ and $0.029\pm 0.006$ respectively; \citealt{Fegleyetal1991,Bainesetal1993}), also presented in \cite{LoddersFegley1994}, were assumed to extend deep into the envelope, and to disentangle the effect of methane on the result, models were constructed with methane abundances at the upper and lower reported error range. While this first-pass assumption may not accurately reflect reality, the main point of this work is to consider the role of water as a possible significant constituent in the envelopes of these planets. We acknowledge that the density contribution of water to the envelope assumed in this work could, in principle, be exchanged with that of ammonia and methane. Assumptions made for the atmosphere will be discussed in the next subsection. Below the assumed water cloud level, we assume for simplicity that the deeper water-rich region is well-mixed.

Moreover, in the ice-rich layer, the included constituents were H$_2$O, CH$_4$, NH$_3$, and varying amounts of H$_2$. The mole fraction of H$_2$ with respect to H$_2$O was defined as described above. Ammonia and methane were assumed to be present in solar proportion of N and C relative to the O of water, in accordance with \cite{Lodders2010}. The caveats of this assumption are discussed in Section \ref{sec:disc}. Finally, the rock core was taken to be comprised of the uniform mixture of SiO$_2$, MgO,  FeS, and FeO assumed by \cite{HubMacfar80}. The mantle and core were assumed to be chemically homogeneous.


\subsection{Derivation of density profiles}

Density profiles were derived beginning at the $1$-bar pressure level \citep{Lindal1992,Lindaletal1987} and integrating to the center of the planet. Specifically, we start at the average $1$-bar level $\bar{R}$ implied by the extrapolated equatorial and polar $1$-bar radii from the Voyager 2 radio occultation data (Table \ref{tab:constr}). 

We assume an adiabatic temperature gradient 

\begin{equation}
T(z)=T_{\text{eff}} \bigg( \frac{P(z)}{P_{e}} \bigg)^{\Gamma},
\label{adiabat}
\end{equation}

where $z$ denotes depth, subscript $e$ refers to values at the outer radius of the given region, and $\Gamma$ is the Gruneisen parameter, $\sim 0.3$ for a solar hydrogen-helium mixture. This is only approximate.

The exception to this rule occurs in the atmosphere, where we account for the effect of water condensation on the temperature, in accordance with \cite{KurosakiIkoma17}, via \cite{Andy1969,Atreya1986,AbeMatsui1988}, by assuming a wet adiabatic temperature gradient:

\begin{equation}
\frac{d \ln T}{d \ln P} = \nabla_{\text{dry}}\frac{1+\frac{x}{1-x}\frac{d \ln p^{*} }{d \ln T}}{1 + \frac{R_{g}}{C_{p}} \frac{x}{1-x} \Big(\frac{d \ln p^{*} }{d \ln T}\Big)^2}
\label{moistadiabat}
\end{equation}

\begin{figure*}
\centering
\includegraphics[width=\textwidth]{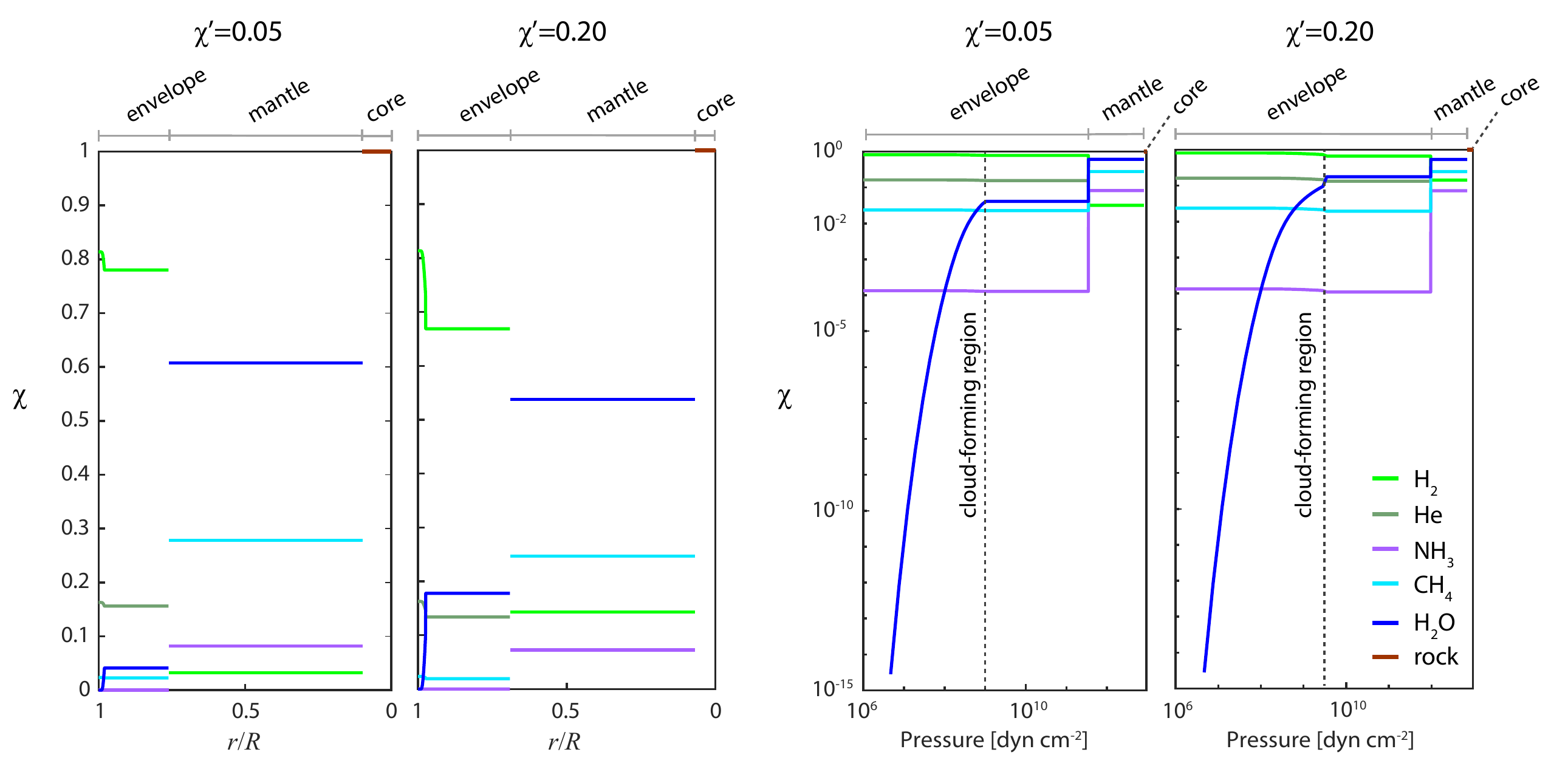}
\caption{Example profiles of abundances of constituents, for $\chi_{\text{env}}'=0.05$ and $0.20$, where $\chi'_{\text{env}} \equiv \chi_{\text{H}_2\text{O}}/(\chi_{\text{H}_2\text{O}}+\chi_{\text{H}_2})$ in the envelope. For these examples, a symmetric coexistence curve was assumed. Examples shown are for Neptune, but a similar scheme was used in Uranus. The left two plots show the mole fractions of constituents as a function of normalized planet radius, while the right two plots show mole fractions in the same models as a function of pressure. Water is taken to be present in the atmosphere at saturation vapor pressure until the cloud-forming region is reached. The cloud-forming region was assumed to occur either when the molar abundance reached the value chosen for the underlying homogeneously mixed region of the envelope (as in the $\chi'=0.05$ case), or when the critical temperature of pure water was reached (as in the $\chi'=0.20$ case), whichever came first. The abundances of homogeneously mixed regions were chosen according to the rationale described in the text. This figure shows models with a separate rock core. For models with no separate rock core, rock is assumed to be uniformly mixed with the mantle material.}
\label{fig:abundprof}
\end{figure*}

where the symbol $\nabla_{\text{dry}}$ refers to the dry adiabatic gradient, while $C_{p}$ is the mean heat capacity, $x_{i}$ the mole fraction of water, and $p_{i}^{*}$ the vapor pressure of water, which was calculated according to \cite{nakajima1992study}. Water was assumed to be present at saturation vapor pressure until the chosen mole fraction of the lower region was reached (Figure \ref{fig:abundprof}). If the critical temperature of pure water (647 K) was reached before the chosen mole fraction of the lower well-mixed region was attained, the 647 K temperature level was taken to be the cloud level, in accordance with \cite{FegleyPrinn1986}. The density contribution of condensed water clouds was neglected, as was the density contribution of any other condensed species. Wet adiabaticity due to ammonia and methane was also ignored--this choice agrees with the finding of \cite{guillot1995condensation} that a moist adiabat is not indicated by the deduced temperature gradient to $\sim 2$ bar in Uranus and $\sim 4$ bar in Neptune. While \cite{guillot1995condensation} also notably finds that condensation of ammonia and methane inhibits convection around the $\sim 1$ to $2$ bar range, the temperature gradient evidently resumes a dry adiabat below these levels, and this and any analogous deeper effect of water condensation is ignored for the purposes of this work.

It should be noted that there is at present no consensus on the correct treatment of the temperature profile in a region where there is a compositional gradient arising from condensation alone. In the "wet adiabat" assumption described above, the latent heat effect can cause the temperature profile to be substantially \textit{colder} than the dry adiabat. On the other hand, a sufficiently large compositional gradient (arising from the strong dependence of vapor pressure on temperature for a condensable such as water) may actually inhibit convection and cause the temperature gradient to be \textit{hotter} (that is, superadiabatic) relative to a dry state \citep{Leconteetal17, friedson2017inhibition}. However, \citep{Lietal20} suggest, based on microwave data from Juno, that convection is not inhibited in Jupiter's atmosphwere and the moist adiabatic assumption is likely to apply.

Hydrostatic equilibrium is also assumed,  taking into account the latitude-averaged centrifugal force (following \citealt{NepBook}):
 \begin{equation}
 \frac{dP}{dr}=-\Big[ \frac{GM(r)}{r^{2}} - \frac{2\omega^{2}r}{3}  \Big]\rho(r),
 \end{equation}
where $P$ denotes pressure, $\rho$ denotes density, $\omega$ the angular velocity, and $M(r)$ the mass contained inside radius $r$. Given the pressure and temperature computed in this manner, and the temperature given by Equation (\ref{adiabat}) at each depth in the planet, the EOS for each constituent is used to determine the resultant density. In the outer few percent of the planet, the ideal gas equation of state (EOS) is assumed. Deeper in the planet we model the density contribution from each constituent with zero-temperature equations of state, with a thermal pressure correction taken into account in the envelope and mantle. The transition between the ideal gas to zero-temperature EOS is assumed several percent of the distance into the planet where the two equations of state cross. 

Before describing our approach to the equations of state at great depth, we need first to make a philosophical point: It is not our goal to have the ``best possible'' descriptions of the constituent materials. We need only to have descriptions that are realistic enough to uncover the differences implied for the planets once our phase diagram assumptions are enforced. This approach is reasonable for Uranus and Neptune because of the large uncertainties in composition and because the inferred differences in the planets are large enough to affect the interpretation of their heat flows and atmospheres. It would be an unreasonable approach for Jupiter or even Saturn where there are very precisely known parameters and there is an obvious need to adopt very precise descriptions of hydrogen in particular. An assessment of how choice of water equation of state affects our results is discussed in Section \ref{sec:results}.

For the zero-temperature equation of states of the assumed constituents, we use the polynomial approximations suggested by \cite{NepBook} and briefly summarized here. The equation of state for molecular hydrogen is taken to be the experimental result of \cite{Maoetal1988} up to pressures of $\sim 8 \times 10^{11}$ dynes cm$^{-2}$, above which we use the approximation for theoretically determined values of \cite{Zharkov78}. For helium, we use the approximation to the equation of state of \cite{Zharkov78}. For water, we use the polynomial approximation of the EOS determined by \cite{Ree76}, as well as the ab initio EOS of \cite{mazevet2019ab}. Due to their smaller expected abundances, the accuracy for $\mathrm{CH_4}$ and $\mathrm{NH_3}$ is less crucial, and again, we use the polynomial approximations given by \cite{NepBook} for the shockwave $\mathrm{CH_4}$ data determined by \cite{Nellisetal1981} as well as the approximation given by \cite{NepBook} for the zero-temperature $\mathrm{CH_4}$ equation of state. Finally, to model the density of rock within the planets, we employ the EOS from \cite{ZharkovTrubitsyn} used by \cite{HubMacfar80} for the mixture of $38\%$ $\mathrm{SiO_2}$, $25\%$ $\mathrm{MgO}$, $25\%$ $\mathrm{FeS}$, and $12\%$ $\mathrm{FeO}$ which we similarly take to approximately constitute ``rock.''

The approximations to the zero-temperature equations of state mentioned above all take on a form
\begin{equation}
P_0=f(\rho),
\end{equation}
where $P_0$ is the electron degeneracy pressure and $f$ is a polynomial function. To account for the effect of thermal pressure in the planets, we apply a thermal correction to the zero-temperature equation of state, accounting for the thermal pressure $P_{t}$:
\begin{equation}
P_0 + P_{t} = f(\rho),
\end{equation}
where, according to Debye theory, $P_{t}$ is approximated to order of magnitude as
\begin{equation}
P_{t}=3n\gamma k T D(\Theta /T)
\end{equation}
where $n$ is the number density of molecules, $\gamma$ is the Gruneisen parameter, $k$ is Boltzmann's constant, $T$ is the temperature, $D$ is the Debye function, and  $\Theta$ is the Debye temperature.

Following \cite{demarcus1958constitution,Peebles64}, we use the linear mixing assumption for the EOS of a mixture of individual constituents:
\begin{equation}
\frac{1}{\rho(P,T)} = \sum_{i} \frac{m_i}{\rho_i(P,T)}
\end{equation}
where $i$ is iterated over all constituents present in the mixture, $\rho(P,T)$ denotes the density of the mixture at a given pressure and temperature, $m_i$ and $\rho_i$ are respectively the mass fraction and density of constituent $i$. Densities of ice mixtures derived using this standard assumption have been found to vary by $\sim 4 \%$ from a real mixture of ices at conditions relevant to the interiors of ice giants \citep{bethkenhagen2017planetary}.


\begin{table}
 \centering
 \begin{threeparttable}[b]
 \caption{Observational constraints used in this work.}
 \label{tab:constr}
 \begin{tabular}{lll}
 \hline
  & Uranus & Neptune \\
  \hline
 total planet mass [kg $\times 10^{25}$]&  $8.68$\tnote{a} & $10.241$\tnote{a}\\
measured equatorial radius $a$ at 1 bar [km] & $25559$\tnote{b}& $24766$\tnote{c} \\
 measured polar radius $b$ at 1 bar [km] &$24973$\tnote{b} & $24342\tnote{c}$ \\
mean planet radius at 1 bar [km]   & $25362$\tnote{d} & $24624$\tnote{d} \\
 present-day effective temperature [K] & 76(2)\tnote{b} & 72(2)\tnote{c} \\
revised solid-body rotation period [s] & $59664$\tnote{e} & $62849$\tnote{e}\\
Voyager rotation period [s] &
$62080$\tnote{b} &
$58000$\tnote{c}\\
quadrupole gravitational harmonic $J_{2} \times 10^{-2}$ & $0.35107(7)$\tnote{f} & $0.35294(45)$\tnote{g}\\ 
octopole gravitational harmonic $J_{4} \times 10^{-4}$ & $-0.342(13)$\tnote{f} & $-0.358(29)$\tnote{g} \\
 \hline
  \end{tabular}
  \begin{tablenotes}
     \item[a] via JPL Horizons. \item[b] \cite{Lindaletal1987}.   
     \item[c] \cite{Lindal1992}. 
     \item[d] $\bar{R}\equiv \sqrt[3]{a^{2}b}$
     \item[e] \cite{Helledetal10}.
     \item[f] \cite{Jacobson14}.
     \item[g] Based on \cite{Jacobson09, Lindal1992} in the same manner as \cite{Helledetal10,Nettelmannetal13}, for a reference radius of the 1-bar pressure level.
  \end{tablenotes}
  \end{threeparttable}
 \end{table}
 
 \begin{figure*}
\centering
\includegraphics[width=0.8\textwidth]{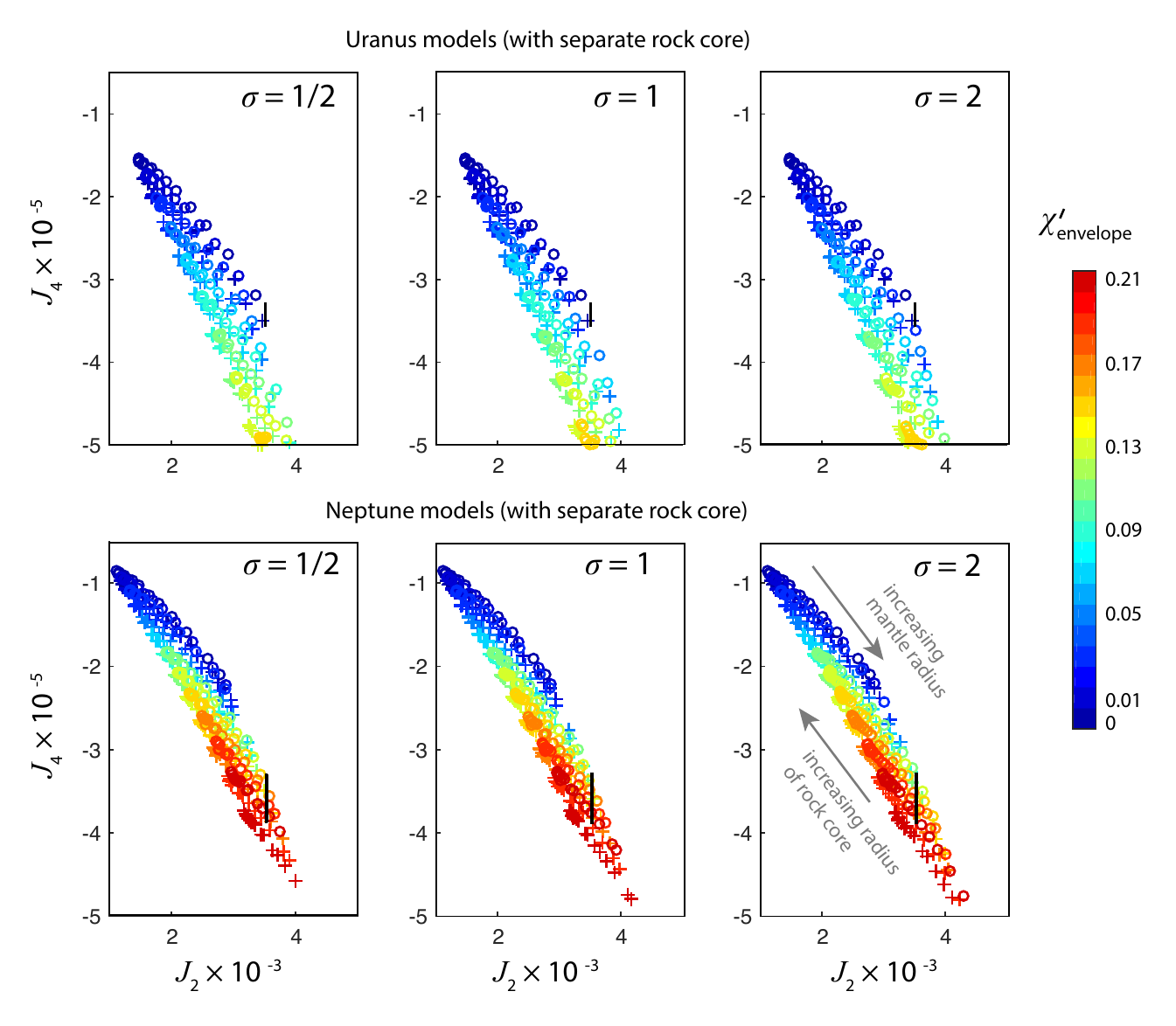}
\caption{Gravitational harmonics derived for three-layer models. Observationally derived harmonics $J_2$ and $J_4$ for Uranus and Neptune \citep{Jacobson14,Jacobson09} are shown as black boxes (the boxes resemble line segments due to sufficiently tight constraints on $J_2$). Colors represent the mole fraction $\chi'_{\text{env}} \equiv \chi_{\text{H}_2\text{O}}/(\chi_{\text{H}_2\text{O}}+\chi_{\text{H}_2})$ in the envelope. Layer compositions were chosen in accordance with the rationale described in Figure \ref{fig:scheme} and in the text. The parameter $\sigma$ describes the assumed asymmetry of the model critical curve and is defined such that $\chi_{\text{env}}
=\sigma \chi_{\text{man}}$, where $\chi_{\text{man}} \equiv \chi_{\text{H}_2}/(\chi_{\text{H}_2\text{O}}+\chi_{\text{H}_2})$, the ratio in the mantle. For every set of layer compositions, a range of models was constructed to satisfy the mean density and radius of the planets, by varying the radius of the ice-rich mantle and rock core, as described in the text. As indicated by the arrows, models toward the lower right have comparatively larger icy mantles and smaller rock cores. The '$\circ$' symbols refer to gravity harmonics derived by taking the spheroid density to be the outer extent of each spheroid, while the '+' symbols refer to the harmonics derived by taking the spheroid density to be that of the outer limit of the adjacent interior spheroid, in accordance with the rationale described in the text.}
\label{fig:rainbow2}
\end{figure*}
 
\begin{figure}
\centering
\includegraphics[width=0.6\textwidth]{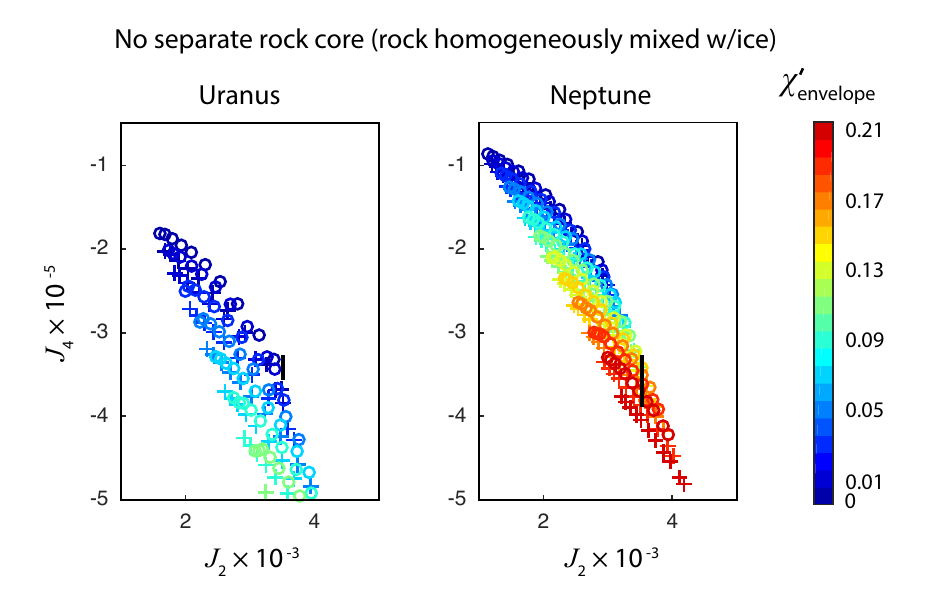}
\caption{Gravitational harmonics for derived models analogous to those shown in Figure \ref{fig:rainbow2} but without a separate rock core (i.e. rock is homogeneously mixed with ice in the mantle).}
\label{fig:rainbow3}
\end{figure}

\begin{figure}
\centering
\includegraphics[width=0.6\textwidth]{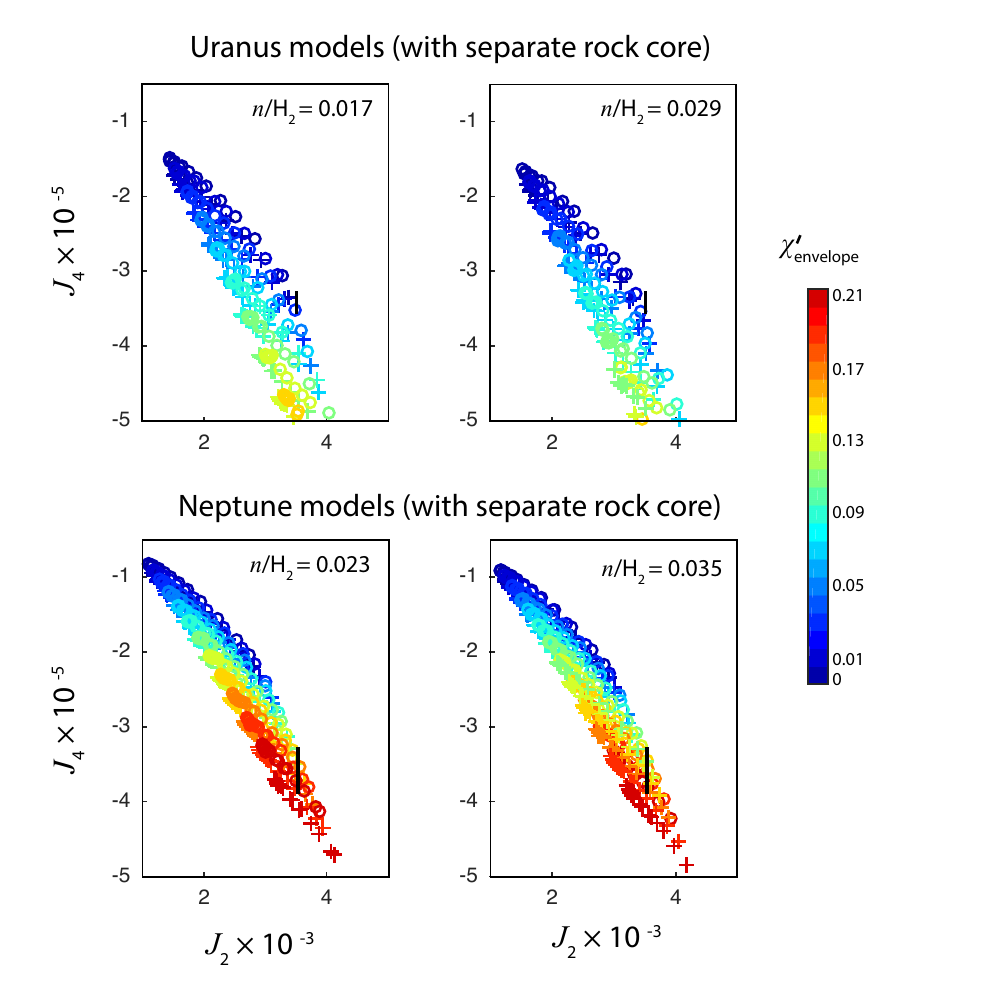}
\caption{Models of Uranus and Neptune assuming a symmetric H$_2$-H$_2$O critical curve ($\sigma=1$ case) with methane abundances $n/\text{H}_2$ relative to hydrogen chosen at the lower and upper observational bounds for each planet (Table \ref{tab:constr}). Colors and observational bounds on the gravity harmonics are defined as in Figure \ref{fig:rainbow2}, and the water EOS of \cite{Ree76} was used.}
\label{fig:methanevariance}
\end{figure}

 \begin{figure}
\centering
\includegraphics[width=0.6\textwidth]{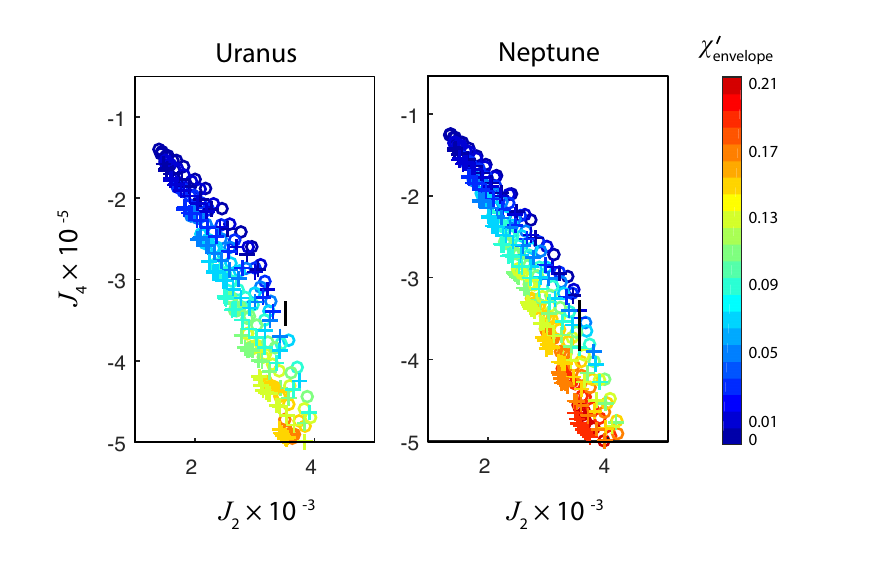}
\caption{Gravitational harmonics $J_2$ and $J_4$ for Uranus and Neptune models, with gravity harmonics calculated using the nominal Voyager rotation rates \citep{Lindaletal1987, Lindal1992}. Colors and observational bounds on the gravity harmonics are defined as in Figure \ref{fig:rainbow2}, and the water EOS of \cite{Ree76} was used.}
\label{fig:VoyRot}
\end{figure}

\begin{figure}[t!]
\centering
\includegraphics[width=0.6\textwidth]{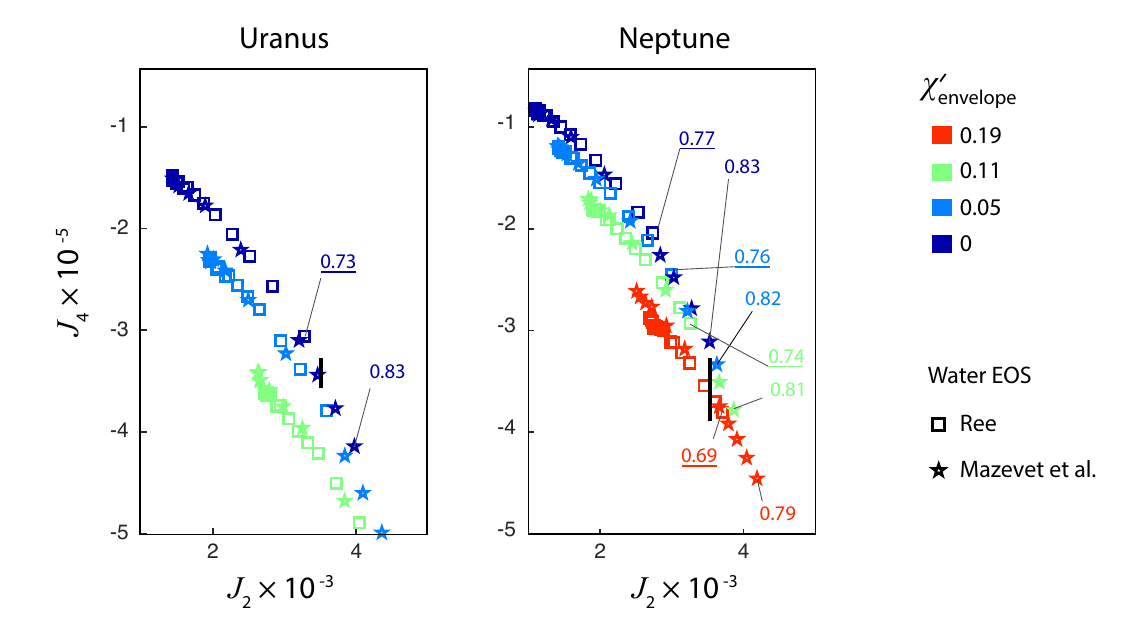}
\caption{Comparison of models assuming different water equations of state. A sample of gravitational harmonics $J_2$ and $J_4$ for Uranus and Neptune models, with gravity harmonics calculated using the \cite{Helledetal10} rotation rates. Equations of state used were \cite{Ree76} (squares) and \cite{mazevet2019ab} (stars). Numbers on plots refer to the location of the upper boundary of the icy mantle, as a ratio of total planet radius. }
\label{fig:EndFig}
\end{figure}

\subsection{Rotation Rates}
\label{ROTRATE}
A crucial input to the determination of the gravity harmonics is the assumed solid-body rotation rates of Uranus and Neptune. In this work, we primarily use the rotation periods determined by \cite{Helledetal10} to minimize the dynamical heights of the 1-bar isobaric surfaces of Uranus and Neptune. While this method produces a plausible result (which is compatible with our observational understanding of Jupiter and Saturn), accurate rotation rates must come from future missions. The implications of assuming the Voyager 2 rotation rates are also discussed in Section \ref{sec:results}. 
 
\subsection{Derivation of model gravity harmonics}
After computing a range of models satisfying the planet mass and mean radius, with varied composition and extent of the layers, we then derived the implied gravity harmonics and compared the result with observations. Due to the presumed presence of a density discontinuity in the outer region of these planets, the Radau-Darwin approximation is not robust \citep{GaoStevenson}. We opt to use the concentric Maclaurin spheroids approach to theory of figures developed by \cite{Hubbard13}, in which the planet is treated as a set of concentric spheroids with homogeneous densities. The shape of the $jth$ spheroid is an equipotential surface found by iteratively solving for the balance of gravitational and rotational potentials. The CMS method was chosen in this work because it allows trivial inclusion of substantial density discontinuities, while permitting density gradients to be modeled with arbitrarily many concentric spheroids. A potential caveat resides in the assumption of solid-body rotation; however, \cite{Kaspietal13} found that the observed gravity fields require that differential rotation extend at most $\sim 1000$ km into Uranus and Neptune, while \cite{Soyueretal2020} find that, to satisfy energy flux constraints, Ohmic dissipation estimates require the wind dynamics to penetrate no deeper than $0.93-0.97R_{\text{U}}$ and $0.95-0.98R_{\text{N}}$.

For both planets, 30 equally spaced spheroids, with an additional spheroid at the location of each transition (i.e. the envelope-mantle transition, and the mantle-core transition in 3-layer models containing a separate rock core), and 15 iterations between the shape and gravity, were found to be more than sufficient to compute the gravitational harmonics to observational precision. One source of error inherent to the CMS method arises from the discretization of density gradients within the planet. Accordingly, two versions of the CMS calculation were computed for each model, which can be considered as providing lower and upper bounds on the gravitational harmonics: one in which the density chosen for each spheroid was the (lower) density occurring at the outer bound of the next interior spheroid, and another calculation with the chosen density being the higher density present at the inner boundary of each spheroidal shell (i.e. the outer bound of the next interior spheroid). The results of these calculations are discussed in the next section. 

\section{Results}\label{sec:results}

Figure \ref{fig:rainbow2} shows the derived gravitational harmonics $J_{2}$ and $J_{4}$ for three-layer models of Uranus and Neptune. For Uranus, the models that best fit the gravity data have a mole fraction in the envelope $\chi_{\text{env}}'\lesssim 0.01$ of water relative to hydrogen. For Neptune, to fit the observed gravitational field, it appears that $\chi_{\text{env}}' \gtrsim 0.10$ is necessary. This result holds for both symmetric and asymmetric assumed critical curves, as seen in Figure \ref{fig:rainbow2}. Evidently, the envelope proportion of water relative to hydrogen dominates the gravity harmonics in these models. The qualitative robustness of this result to asymmetry in the phase diagram suggests that, for interior models of ice giants which take into account hydrogen-water mixing constraints, it is necessary to include a substantial proportion of metals in Neptune's envelope, compared to Uranus, which must be more centrally condensed. Evidently, this appears to point to the nature of a dichotomy between these two superficially similar planets, which may be related to their disparate heatflows, as we will discuss further in teh next section. In particular, if hydrogen-water phase separation, as hypothesized in this work, does hold--and if Neptune is indeed in a less demixed state than Uranus--we find that an explanation naturally arises for the large heat flow of Neptune relative to Uranus.

Next, we consider the dependence of this result on the assumption of rock-ice mixing properties. While, as discussed in the prior section, the gravitational harmonics dictate that a transition must take place between a lighter, hydrogen-rich region to an intrinsically denser region, it is also warranted to quantify the possible effect of a separate rock core on the gravity field. As gravitational harmonics generally probe the density structure at levels above $\sim 0.5 R_{\text{planet}}$, it might be expected that the inclusion of a separate rock core versus homogeneously mixed rock might not make a substantial difference to our result. To check that this expectation holds, Figure \ref{fig:rainbow3} shows the gravitational harmonics for the associated two-layer models--including rock mixed in the mantle, rather than in a separate core, as described in the previous section--for the case of a symmetric coexistence curve. In the two-layer case, the mantles--which include uniformly mixed rock in the same quantity originally relegated to the separate core in the three-layer models--extend a fraction of a percent lower per total planet radius due to self-compression; however, the effect on the density profile above $r\sim 0.5 R_p$, the region predominantly sampled by the $J_2$ and $J_4$ gravity harmonics, is typically negligible. Indeed, although slight variation from the three-layer case is observed, Figure \ref{fig:rainbow3} shows that the resulting gravity harmonics are qualitatively similar to the result for three-layer models.

Moreover, we took into account the effect on our result of the uncertainty for the abundance of methane in the envelopes of Uranus and Neptune. Models with methane abundances chosen at the upper and lower observational bounds for each planet (Table \ref{tab:constr}) are shown in Figure \ref{fig:methanevariance}, again for the case of symmetric H$_2$-H$_2$O coexistence curve. Although the model gravity harmonics can vary by as much as several percent within the observational error range for methane abunance, the result is qualitatively similar at the upper and lower bounds of the measured concentrations.

As mentioned in Section \ref{ROTRATE}, updated estimates \citep{Helledetal10} of the rotation rates of Uranus and Neptune differ from the Voyager rotation rates by $\sim 1$ hour. The model gravitational harmonics for Uranus and Neptune that result from considering the original Voyager rotation rates are shown in Figure \ref{fig:VoyRot}. Notably, assumption of the Voyager rotation rates does qualitatively change the layer compositions capable of producing the observed gravity harmonics--in particular for Neptune, whose gravity harmonics can then be satisfied with a lower-metallicity envelope. While higher metallicities in Neptune's envelope are still permitted, the Voyager rotation rates are associated with some breakdown in the differences between the envelope compositions allowed in Uranus versus Neptune. This result highlights the utmost importance of future missions in obtaining reliable rotation period estimates. 

Moreover, Figure \ref{fig:EndFig} shows a comparison of models produced assuming two equations of state for water, \cite{Ree76} and \cite{mazevet2019ab}. A smaller density of water at the relevant conditions, implied by the \cite{mazevet2019ab} equation of state, allows the mantle to extend further from the planet center while still allowing the model to satisfy the total planet mass, translating to higher permitted $J_2$ of models with low envelope metallicity. In a manner similar to the results assuming Voyager rotation rates shown in Figure \ref{fig:VoyRot}, Neptune models with high envelope metallicity ($\chi_{\text{env}}' \gtrsim 0.01$) are permitted by both equations of state, while a comparison of Uranus models to the gravity data appear to suggest a low-metallicity envelope ($\chi_{\text{env}}' \lesssim 0.01$) may be necessary to explain the observed gravitational field.

\begin{figure}
\centering
\includegraphics[width=0.7\textwidth]{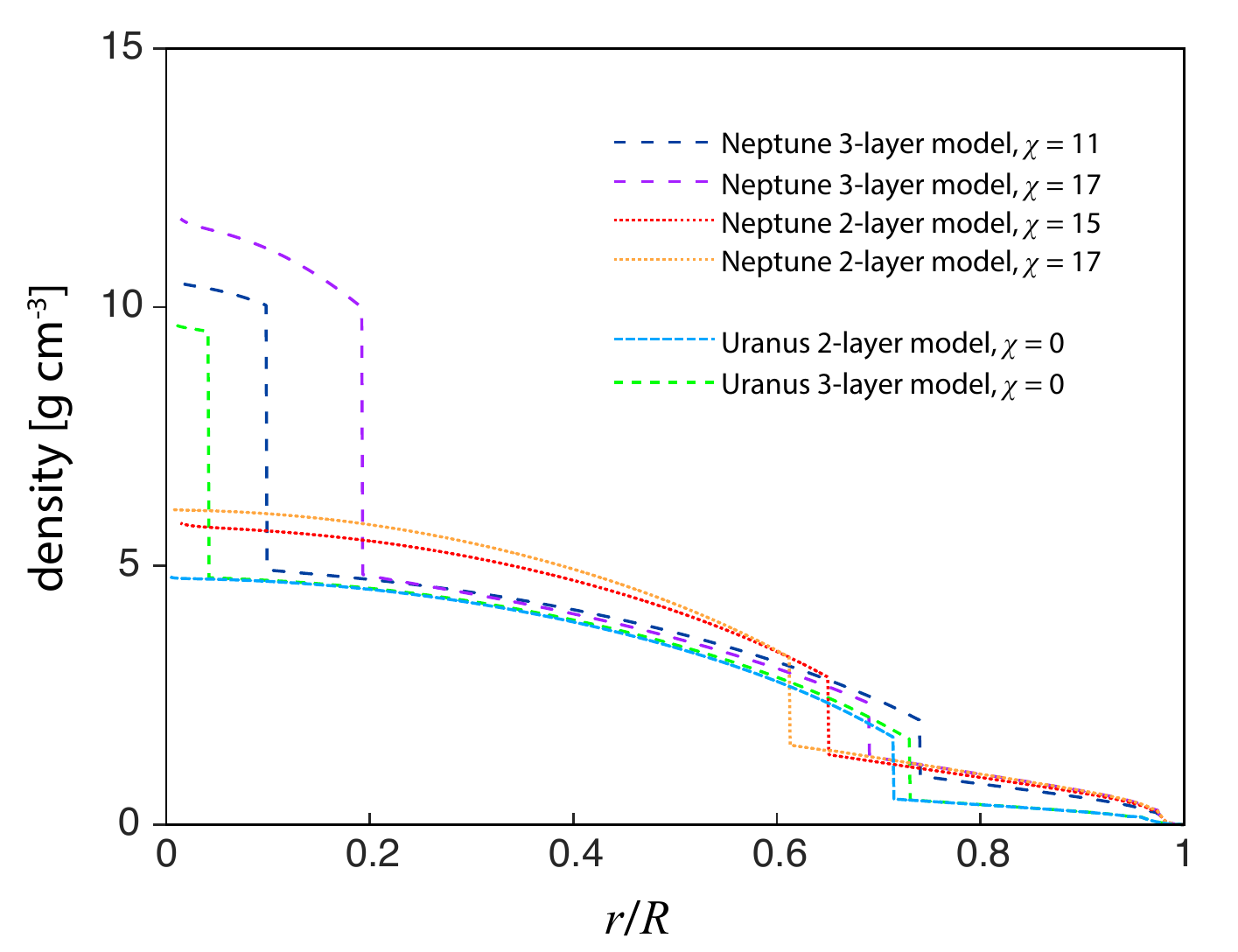}
\caption{Examples of derived two- and three-layer density profiles for Uranus and Neptune which were found to approximately reproduce observed $J_2$ and $J_4$.}
\label{fig:profpic}
\end{figure}

Examples of density profiles of two- and three-layer models of Uranus and Neptune producing a close fit to the observed gravitational harmonics with the assumed solid-body rotation, are shown in Figure \ref{fig:profpic}. While these models generally appear quite similar to many other published models of these planets that assume homogeneous layers, the key difference in this case is that our novel mixing constraint (Figure \ref{fig:scheme}) has been imposed for the layer compositions of the envelope and mantle, removing a degree of freedom compared to traditional layered models.

We now turn to further discussion of these results. In particular, we consider in some detail the effect our hypothesized demixing scenario--and associated gravitational potential energy release--could have on he heatflow. We also relate these results to the observed atmospheric abundances of methane and ammonia.

\section{Discussion}\label{sec:disc}

Our finding that Neptune, in the model framework discussed in this work, would require substantial metallicity of its outer layer to account for the observed gravity harmonics, agrees with results of other layered models (e.g. \cite{Nettelmannetal13}) showing that Neptune can have substantial metal enrichment in its envelope compared to Uranus in models that satisfy the planets' gravity fields and updated rotation rates from \cite{Helledetal10}. From these gravity and rotation data, Neptune is potentially expected to be less centrally condensed than Uranus--a difference that has been suggested by \cite{reinhardt2020bifurcation} to result from different giant impact histories of Uranus and Neptune. This work presents another, potentially related hypothesis for this dichotomy--namely, that both planets might contain demixed hydrogen and water as major constituents, and that Neptune is currently in a considerably less-demixed state than Uranus. To test this new hypothesis, it will be critical to resolve the disagreement between the experimental work of \cite{Bali13} and the ab initio study of \cite{SoubMil} addressing the question of whether hydrogen and water are miscible in the deep interiors of ice giants. Moreover, if it is found that hydrogen-water immiscibility is in fact expected in Uranus and Neptune, the specific nature of the hydrogen-water critical and coexistence curves will be of utmost importance in constraining the interior states of these planets. As will now be discussed, the properties of the hydrogen-water system may also be crucial for understanding the disparate heat flows of Uranus and Neptune.

In the model framework presented in this work, as Neptune cools, the equilibrium mole fraction $\chi_{\text{env}}'$ of water in the envelope should currently be decreasing with time. This process is expected to be associated with a change in gravitational potential energy in the planet. Our finding that Uranus may have considerably less water in its envelope than Neptune suggests that, unlike Neptune, hydrogen-water demixing in Uranus could be at or near completion, and no longer contributing to the heatflow. Accordingly, it may be worth considering the role of present-day gravitational energy release due to present-day hydrogen-water demixing in Neptune but not Uranus, as a potential major source of the observed heatflow in Neptune.

\begin{figure}
\centering
\includegraphics[width=0.8\textwidth]{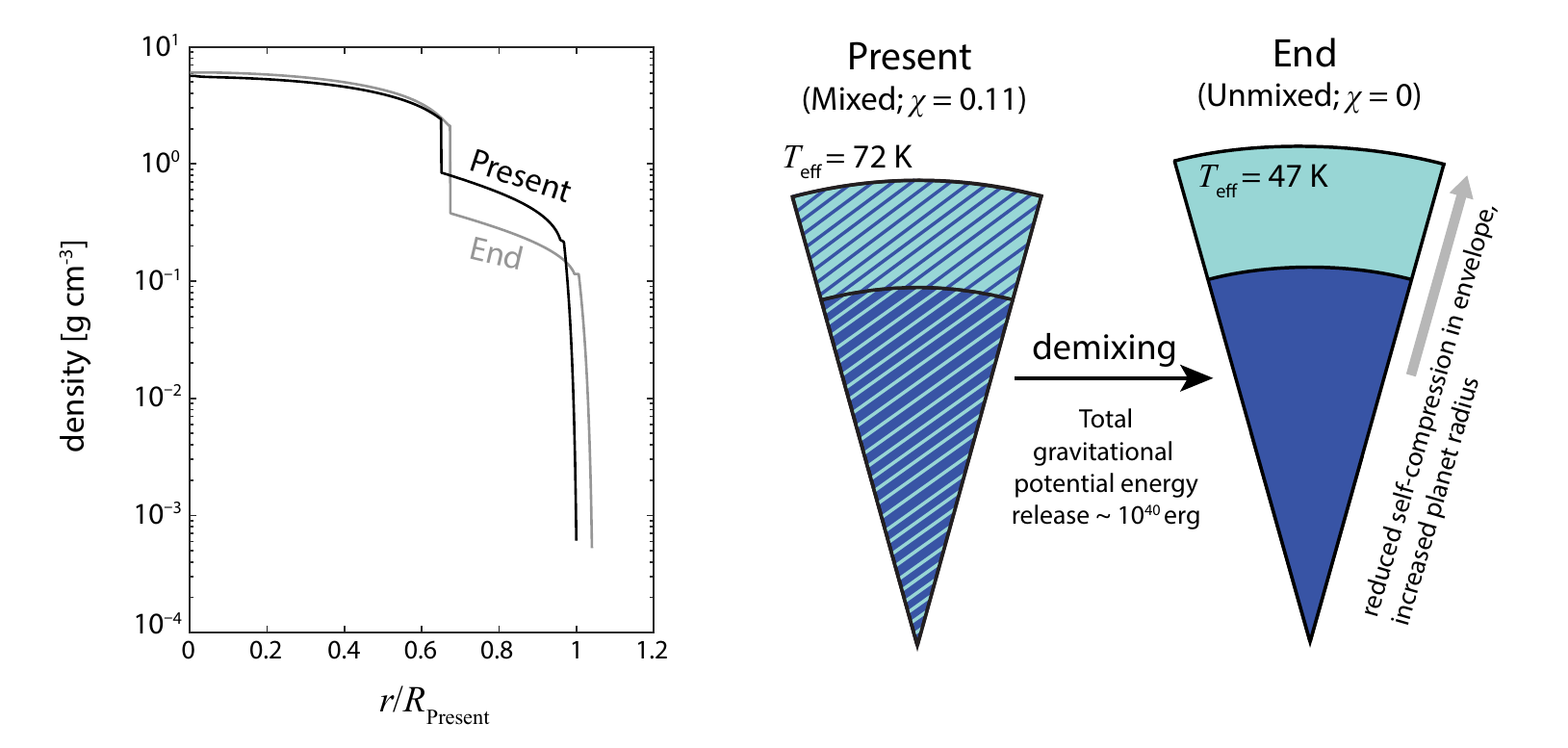}
\caption{The estimated gravitational potential energy difference between Neptune's inferred present-day mixing state with the unmixed future evolutionary endmember is $\sim 10^{40}$ erg, sufficient to supply Neptune's present-day observed heat flux, $2\times10^{22}$ erg \citep{podolak1991models,PearlCon1991}, for roughly $10$ solar system lifetimes. }
\label{fig:best}
\end{figure}

To test whether it is feasible for hydrogen-water separation to account for the present-day observed heatflow of Neptune, the available gravitational energy release from this process was estimated as follows. An approximate model of Neptune's interior was constructed by taking the masses of hydrogen and water present in the original best-fit $\chi=0.11$ Neptune model (assuming a symmetric critical curve), and then recalculating an associated simplified present-day two-layer model, containing only hydrogen and water in the original proportions.\footnote{The other constituents are neglected for the purposes of this first-pass model, but it is expected that the nonpolar species (e.g. He and CH$_4$) would be incorporated into the hydrogen, and the polar species (e.g. NH$_3$) with the water. Improved constraints on abundances of polar and nonpolar constituents relative to solar in the envelopes of these planets might eventually shed light on whether this mechanism is present, and may potentially even serve as an indirect test of hydrogen-water miscibility in these planets.} As in the previous, more detailed models, adiabaticity of the interior was assumed, with $T_{\text{eff}}$ of the present-day model assumed to be the same value as before ($72$ K). The resulting present-day model of Neptune has a total radius $\sim 92\%$ that of the original, the same order of magnitude. Next, a model of Neptune's future demixed end state was constructed, with an envelope of pure hydrogen and a mantle of pure water, and the total masses of each constituent held fixed compared to the present-day model (Figure \ref{fig:best}). The effective temperature of the end-state model was derived by equating the power absorbed to the power radiated,

\begin{equation}
T_{\text{eff}}=\sqrt[4]{\frac{L(1-a)}{16 \pi \sigma D^2}} \sim 47 \text{ K},
\end{equation}

where the solar luminosity $L$ and bond albedo of Neptune $a$ \citep{PearlCon1991} are assumed not to deviate from present-day values, $\sigma$ is the Stefan-Boltzmann constant, and $D$ is orbital distance from the sun.

Due to reduced self-compression in the envelope, the total planetary radius in this model was found to increase by $\sim 4\%$ relative to the initial present-day simplified model. Comparing the total gravitational potential energy in the two models, we find that the available total gravitational potential energy release from hydrogen-water demixing in Neptune is $\sim 10^{40}$ erg, sufficient to supply Neptune's present-day heat flux of $\sim 3 \times10^{22}$ erg/s \citep{podolak1991models,PearlCon1991} for roughly ten solar system lifetimes. This rough estimate of total available energy appears to suggest that gravitational potential energy release from hydrogen-water demixing could plausibly supply Neptune's entire present-day heat flux. 

It is natural to compare the proposed process of hydrogen-water demixing in Uranus and Neptune with the well-known mechanism of helium rainout expected to occur in gas giants due to the immiscibility of helium in metallic hydrogen. This latter process has been proposed \citep{stevenson1977phase,stevenson1977dynamics} to account for the luminosity excess of Saturn, as well as atmospheric depletion of helium in Saturn \citep{Stevenson1980,Conrathetal1984}. But while both helium rainout and the currently proposed mechanism of continued hydrogen-water demixing invoke potential energy release of separating constituents as a contributor to the planet's luminosity, the two processes are not perfectly analogous. A key difference exists--more specifically, helium rainout, as it is generally discussed, occurs when the cooling giant planet's adiabat crosses into a regime of immiscibility for helium and metallic hydrogen, causing the helium to become immiscible and rain out (bringing dissolved noble gases with it). In contrast, with the presently discussed mechanism of continued hydrogen-water demixing, it is assumed that hydrogen and water are \textit{already} immiscible and separated into two phases in ice giant interiors. The gravitational potential energy release in Uranus in Neptune is instead proposed to be due to variation of the equilibrium compositions of the already-separated phases, as the planet cools and approaches a state where hydrogen and water are completely demixed. The difference arises primarily because of the much lower pressure of relevance for the hydrogen-water system, though of course it is contingent on unknown aspects of the phase diagram at high pressure and can therefore only be viewed as a hypothesis.

We now proceed with a somewhat more detailed consideration of the effect of hydrogen-water demixing on Neptune's cooling rate, adapting the precedent set forth for helium rain in gas giants by \cite{stevenson1977dynamics}, to the hypothesized phase separation addressed at present for ice giants. As previously described in this section, the planet is treated as consisting of a compositionally homogeneous envelope comprised of the hydrogen-dominant phase, with a compositionally homogeneous mantle comprised of the water-dominant phase. The envelope is assumed to be adiabatic, in agreement with magnetic field observations. The phase separation of hydrogen and water dictates the existence of two simultaneous, opposing gravitational effects: the downward redistribution of water from the envelope to the mantle, and the upward redistribution of hydrogen from the mantle to the envelope. Assuming a critical curve roughly symmetric in mole fraction, as we have done in this work, the proportion of redistributed hydrogen is about an order of magnitude less than the distributed water by mass. Therefore, because the average vertical displacement is similar for both, the gravitational effect of rising hydrogen is neglected for the purposes of this work, and we focus on the effect of water redistributing from the envelope to the mantle. The mass of redistributed water necessary to change the mole fraction of water in the mantle from $x$ to $x+dx$ is approximately

\begin{equation}
M_{\text{H}_2\text{O}} \approx \frac{9 dx M_{\text{man}} }{(1-x)(1+8x)}
\end{equation}

where $M_{\text{man}}$ is the total mass of the mantle. This relation can be obtained by equating the water mass ratio of the mantle expressed in terms of $M_{\text{man}}$,  $M_{\text{H}_2\text{O}}$, and $x$, with the expression of this ratio in terms of $x$ and $dx$, and treating $x+dx \sim x$. The gravitational energy release is approximately

\begin{equation}
E_{\text{grav}} \approx M_{\text{H}_2\text{O}} gH
\end{equation}

where $g \sim 1400$ cm s$^{-2}$ is roughly the average gravitational acceleration inside the mantle and envelope, and $H\sim1.2\times10^9$ is the approximate vertical height between the centers of mass of the two layers. Moreover, the temperature $T_b$ at the boundary between the envelope and mantle is related to the mole fraction $x$ of water in the mantle by the hydrogen-water coexistence curve. Unfortunately, this curve is not yet known, so we resort to using our model critical curve as described earlier for a two-component regular solution, stretched so that the value $T_b\sim3800$ K from our good fit $\chi \sim 0.11$ adiabatic Neptune model corresponds to an equilibrium saturation mol fraction of $\chi\sim0.11$. The value of the critical temperature derived in this manner is then $T_c \sim 7000 K$; comparing this value to Figure 2, this derived critical temperature is indeed potentially suggestive of the aforementioned critical curve turnover expected in these planets.

The parameter $T_0 \equiv (dT_b/dx)_{\text{coex}}$ is the temperature change of the envelope-mantle boundary as dictated by the coexistence curve. Specifically, $T_0$ can be viewed as the tangent slope to the scaled model coexistence curve (i.e., scaled such that $T_c \sim 7000$ K) at the present-day inferred mole fraction of $\chi\sim0.11$ (i.e. $x=0.89$ for the main constituent). While we again emphasize that the critical curve remains unknown, we employ our model critical curve for now, and estimate $T_0 \sim 8000$ K.

Then, treating the mantle as adiabatic\footnote{As discussed earlier in Section \ref{sec:intro}, the mantle may well not be adiabatic, but it is difficult to account for the observed magnetic fields if the outer $\sim 10 \%$ is not adiabatic. (It is convenient for models of Uranus and Neptune that the thermal correction to the zero-temperature equations of state is negligible in the mantles, particularly in the deeper regions of the mantles where convection may be inhibited. The temperature gradient is therefore of critical importance only for determining the density profiles of the hydrogen-rich outer envelopes, and our ``fully adiabatic'' models of these planets are fully compatible with a scenario of stable stratification in the deep mantle.)} If heat transport is inhibited between an outer convecting region and inner stratified region, the thermal energy increase of the mantle may be less, making the ratio $E_{\text{th}}/E_{\text{grav}}$ (which we are about to discuss in the text) even smaller., the thermal energy increase $E_{\text{th}}$ of the mantle can be expressed \citep{stevenson1977dynamics} as

\begin{equation}
E_{\text{th}} \equiv AC_{v}T_{0}dxM_{\text{man}}
\end{equation}

where $A$ is the ratio of typical internal temperature to $T_b$ and is of order $2$, $C_v \sim 2 \times 10^7$ erg g$^{-1}$ K$^{-1}$ is the specific heat of the mantle, and $M_{\text{man}}$ is the mass of the mantle. (Note that this treatment assumes the mantle is where the thermal energy is deposited; as the extrapolated curvature of the critical curve relative to the adiabat suggests (Figure \ref{fig:RAT}), the envelope may be undersaturated in the regions above the envelope-mantle boundary. It is the critical temperature at this boundary which we assume dictates the coexisting phase compositions of the envelope and mantle.) With the above parameterization, the ratio of thermal energy increase to gravitational energy release due to demixing water in Neptune can be approximated as 

\begin{equation}
\frac{E_{\text{th}}}{E_{\text{grav}}} \approx \frac{AC_{v}T_{0}}{9gH} \approx 0.02
\end{equation}

 This ratio is an upper bound, as discussed in the recent footnote. In a similar vein as \cite{stevenson1977dynamics}, the small value of this ratio suggests most of the released gravitational energy is radiated. Therefore, we will proceed to estimate the change in effective temperature with time. We treat the envelope as adiabatic, such that

\begin{equation}
\frac{T_b}{T_e} \equiv \Big( \frac{P_b}{P_e} \Big) ^{\Gamma}
\end{equation}

where $T_e$ is the effective temperature, $P_e$ is the corresponding pressure, and $\Gamma \sim 0.3$ is the adiabatic index. Because the water content of the outer atmosphere is negligible due to the low vapor pressure of water at the relevant temperatures (Figure \ref{fig:abundprof}), demixing of water from the envelope should not affect $P_e$, so we treat it as roughly constant. Moreover, although $P_b$ can decrease by a factor of several over the entire lifetime of the planet, due to the demixing of the water from the envelope and the associated diminishment of overpressure, $P_b$ and $\Gamma$ are treated as roughly constant in a time window surrounding the present day. Therefore $d \ln T_b/dt\approx d \ln T_e/dt$, and in a similar manner as \cite{stevenson1977dynamics}, the gravitational energy release over time can be expressed as

\begin{equation}
Q_{\text{grav}} \approx \frac{9 M_{\text{man}}}{(1-x)(1+8x)T_0} \frac{dT_b}{dt} gH
\end{equation}

and if $Q_{\text{grav}}$ is equated with the present-day observed heat flux $\sim 3 \times 10^{22}$ erg s$^{-1}$ of Neptune, for $M_{\text{man}} \sim 8 \times 10^{28}$ g, $x \sim 0.89$, $T_0\sim8000$ K, $g \sim 1400$ g cm$^{-2}$, and $H \sim 1.2 \times 10^9$ cm, then, accounting for the logarithmic temperature relation between $T_b$ and $T_e$, the cooling rate is estimated as $dT_e/dt \sim 0.1$ K Gyr$^{-1}$. Comparing this to the estimated present-day cooling rate of roughly $\sim 2$ K Gyr$^{-1}$ found for standard adiabatic cooling models (e.g. \cite{HubMacfar80}) that do not account for gravitational energy release of phase separation, it is evident that the hypothesized demixing could indeed significantly prolong cooling.

\subsection{Relationship of model to atmospheric abundances} 

We next turn to a discussion of our model as it may relate to the observed abundances of minor constituents in the atmospheres of Uranus and Neptune. In the partially analogous case of helium rainout in gas giants discussed above, nonpolar neon is proposed to dissolve into the helium rain droplets, leading to atmospheric depletion of neon in gas giants when helium rainout occurs \citep{RoulstonStevenson1995}. The depletion of neon by an order of magnitude relative to solar observed by Galileo in Jupiter's atmosphere \citep{Niemannetal1996} has been interpreted as possible evidence for the commencement of helium rainout in Jupiter \citep{WilsonMilitzer10}. In a somewhat analogous manner, in the potential framework for understanding Uranus and Neptune that has been discussed in this work, polar constituents (e.g. ammonia) may be expected to partition preferentially into the water-rich phase, while nonpolar constituents (e.g. methane) may be expected to partition preferentially into the hydrogen-rich phase. This partitioning will be progressive and grow as the demixing proceeds. Hence, if our framework of immiscibility is correct and if it is true that the demixing of Uranus is further advanced than Neptune, then there may exist the expectation of ammonia depletion in the atmosphere of Uranus relative to Neptune. In a related sense, \cite{dePateretal1991} find that the Voyager radio occultation data \citep{Lindaletal1990} are best in agreement with the presence of an ammonia ice cloud at the $\sim 5$-bar level in Neptune. While the radio occultation data did not probe deep enough on Uranus to make an analogous determination, ground-based observations \citep{Gulkis1978} seem to indicate emission at short cm wavelengths from below the analogous level in Uranus, indicating that such an ammonia cloud might potentially be absent on Uranus. The generally accepted explanation for the atmospheric ammonia discrepancy between the two planets invokes possible differences in atmospheric convection, which might allow some ammonia to bypass depleting reactions with H$_2$S in Neptune but not Uranus. While this explanation remains entirely plausible, if atmospheric depletion of ammonia in Uranus relative to Neptune does exist, it could also potentially be explained as possible evidence for the more advanced demixing of the atmosphere of Uranus relative to Neptune. Hence, this work motivates atmospheric observations to better inform the interiors of Uranus and Neptune.

In an analogous fashion to ammonia in the envelope, methane could potentially demix from the mantle phase, possibly leading to enrichment over time of methane in the atmosphere of Uranus compared to Neptune. The atmospheric methane abundances of Uranus and Neptune, ($n/\text{H}_2 = 0.023 \pm 0.006$ and $0.029 \pm 0.006$ respectively) from Voyager spectroscopic measurements (\citealt{Fegleyetal1991,Bainesetal1993} via \citealt{LoddersFegley1994}) do not indicate a clear discrepancy in atmospheric methane between the two planets, although the measurements are consistent with Uranus having up to $26\%$ more methane in its atmosphere than Neptune. More work is needed to understand the mixing properties of methane and water at conditions relative to the interiors of these planets, to determine whether the methane abundances are consistent with the hypothesized relative demixing states of Uranus and Neptune.

\subsection{Potential caveats of models} 

In addition to the points already addressed in Section \ref{sec:results}, there exist several caveats inherent to the theoretical framework used in this work, which are now discussed.

\subsubsection{Superionicity of water} 

The relevance of H$_2$-H$_2$O mixing properties are expected to diminish at depths in the planet where the hydrogen and water molecules become some other configuration of hydrogen and oxygen atoms. In particular, the occurrence of superioinic ice phases in the lower regions of the inferred ice mantle (e.g. \citealt{WWM13, bethkenhagen2015superionic}) is expected to affect the relevant species interactions at those depths, as well as the densities. These effects are not relevant at the shallower depths ($\sim 0.7 R_{\text{planet}}$) at which the transition from a hydrogen-dominated envelope to heavier materials must occur to satisfy the planets' gravity fields. In our assumed model framework of hydrogen-water immiscibility, if it is assumed this density change is due to a phase transition between coexisting hydrogen- and water-rich phases, then the coexistence curve at the $P-T$ conditions of this transition zone would be expected to govern the layer compositions. Accordingly, the superionic behavior of ices at greater pressures would be expected to be irrelevant to the compositions of the envelope and upper mantle. It may be relevant to the presence or absence of a separate ``rock'' component. As shown by Figures \ref{fig:rainbow2} and \ref{fig:rainbow3}, in our models, it is the composition of the envelope that appears to have the predominant effect on the gravitational harmonics.

\subsubsection{Assumed ice ratios} The ratios of ammonia, methane, and water chosen for the envelopes and mantles of our models may not correctly reflect what is really present. The compositional degeneracy inherent to these intermediate-mass planets ensures that the ice-like density inferred for their deep interiors could be satisfied by numerous combinations of ices, rocks, and hydrogen. While we assume the interiors of both planets contain mantles with well-mixed ices in the same fixed ratio, alternatively, the ratios of interior ices could be different in Uranus and Neptune due to different formation conditions. As Figure \ref{fig:rainbow2} shows, varying the hydrogen content of the mantle by a factor of two relative to water makes little difference to the model gravitational harmonics, even at significant hydrogen mole fractions of $\chi'_{\text{man}}\sim 20 \%$. This invariance of the gravitational harmonics to $\chi'_{\text{man}}$ appears to suggest that the specific relative ratios of ammonia, water, and methane in the mantle would not affect the result that Neptune models favor a substantial ($\gtrsim 10 \%$) mole fraction of metals in the envelope (in this work presumed to be water) and that Uranus requires a considerably smaller fraction ($\lesssim 1 \%$). Given that ammonia--and, in particular, methane--are less dense than water, enrichment of these ices relative to the assumed solar values based on C, N, and O abundances would allow further extension of the outer boundary of the ice-rich mantle from the planet center, and would therefore be expected to alter the resulting range of model gravitational harmonics in a manner analogous to that observed when an intrinsically lighter water equation of state is assumed (Figure \ref{fig:EndFig}). Notably, with an intrinsically lighter ice mixture, additional rock could also be included in models while still satisfying the total planet mass. However, enrichment of these more volatile ices relative to water might be expected to be less likely than enrichment of \textit{water} relative to the other ices, in which case the range of mantle extents satisfying the total planet mass would be more limited. A detailed treatment of ice ratios in the mantles of these planets is a possible worthwhile area of future study, particularly as equations of state of ices become better-resolved and the prospect of an ice giant mission is considered.

\subsubsection{Adiabatic assumption} As mentioned earlier, the assumption of a fully adiabatic interior may not hold for Uranus and Neptune, especially as inhibited convection has often been invoked to possibly explain the low heat flux of Uranus (e.g. \citealt{podolak1991models, Net16, LeconteChab12, Podolak19}). For the purposes of this work, the temperature gradient primarily affects the density distributions inferred from the equations of state. However, the presence of a dynamo in both Uranus and Neptune is thought to require convection in at least the outer $\sim 20$ percent of these planets \citep{StanBlox04, StanBlox06}. Below these depths, the thermal pressure correction is of minimal consequence to the equations of state. Above these depths--if a non-convective region exists--the resulting effect on temperatures, in a region where thermal pressure is important, could potentially have a significant effect on the inferred composition \citep{PodHelSchu19}.

\section{Conclusions}\label{sec:conc}

It is standard practice to model the interiors of Uranus and Neptune as consisting of discrete, compositionally homogeneous layers. However, as discussed in Section \ref{sec:intro}, the choice of specific layer compositions--and the assumption of a discrete boundary between these layers--have not previously received rigorous physical justification. Accordingly, this work presents the first thermodynamically justified models of ice giant interiors, in which the layers and their compositions are based on the inference of hydrogen-water immiscibility in these planets. While the true mixing properties of hydrogen and water remain to be resolved at conditions relevant to the interiors of these planets, the presence of immiscible hydrogen and water in Uranus and Neptune would offer physical justification for a sharp compositional transition from the outer hydrogen-rich envelope to the deeper region of heavier constituents, a transition that is known from the gravity data to be necessary. If this transition is not discontinuous and a substantial density gradient instead exists in the outer $\sim 30$ percent of these planets, it is then challenging for models to produce convective flows sufficient to generate the observed magnetic fields.

Having produced models with the novel thermodynamic constraints applied, we have found that, to satisfy the mean planet density and measured gravitational field, Neptune may contain a substantial portion of water in its hydrogen-dominated envelope, potentially as large as $\chi_{\text{env}}' \gtrsim 0.10$. In contrast, models capable of satisfying these constraints for Uranus seem to require a much smaller metallicity in the outer shell, $\chi_{\text{env}}'\lesssim 0.01$. As discussed in section \ref{sec:disc}, the inferred continued demixing of hydrogen and water in Neptune but not Uranus could possibly account for the disparity in heatflow between the planets. This disparity has long been a challenge to explain, and most of the focus has been on Uranus' unexpected lack of heatflow, rather than Neptune's significant heatflow. However, if the lack of intrinsic heatflow from Uranus is due to inhibited convection in the deep interior, it then becomes a challenge to explain why Neptune is different from Uranus in its heat output. Hence, the mechanism proposed in this work--of gravitational potential energy release due to present-day hydrogen-water demixing in Neptune but not Uranus--may offer a potential means to explain how Neptune could produce the observed heat flow, if its deep interior convection is indeed inhibited in a manner similar to Uranus.

Importantly to the understanding of these planets' interior dynamics, if further experimental and/or theoretical work ultimately were to confirm that hydrogen and water are immiscible at conditions of ice giant interiors beyond $\sim 3$ GPa, the hydrogen-water coexistence curve could then provide a novel means to inform the compositions of the deep, intermediate-mass mantles. Accordingly, the thermodynamic rationale presented in this work provides a tentative approach to potentially infer the internal compositions of Uranus and Neptune. As gravity and magnetic field data cannot provide unique solutions to the deep interior compositions of these planets, it may ultimately be necessary to turn to chemical reasoning to resolve their bulk compositional degeneracies--and accordingly, to guide our understanding of their place in solar system formation.
\\

\textbf{\textit{Acknowledgements.}} The authors thank Yayaati Chachan, Imke de Pater, Ravit Helled and the attendees of the 2020 Bern ISSI Ice Giants meeting, for thoughtful discussions. We thank the anonymous reviewers for their careful review and thoughtful comments which guided the direction of this manuscript. 

\bibliography{UNrefs}{}
\bibliographystyle{aasjournal}

\end{document}